\documentclass{article}
\usepackage{amsmath,amssymb,amsfonts}

\newcommand{\beq}{\begin{equation}}
\newcommand{\eeq}{\end{equation}}
\newcommand{\ben}{\begin{eqnarray}}
\newcommand{\een}{\end{eqnarray}}
\newcommand{\be}{\begin{eqnarray*}}
\newcommand{\ee}{\end{eqnarray*}}

\newcounter{eqalph}
\newcounter{equationa}

\let\ssection=\section
\renewcommand{\section}{\setcounter{equation}{0}\ssection}

\newcounter{example}[section]
\newcounter{remark}[section]
\newcounter{theorem}[section]
\newcounter{proposition}[section]
\newcounter{lemma}[section]
\newcounter{corollary}[section]
\newcounter{definition}[section]

\def\theremark{\arabic{section}.\arabic{remark}}
\def\thetheorem{\arabic{section}.\arabic{theorem}}

\def\thedefinition{\arabic{section}.\arabic{definition}}

\newenvironment{example}{\refstepcounter{remark}\medskip\noindent{\bf
Example \theremark:} }{$\Box$ \medskip}
\newenvironment{remark}{\refstepcounter{remark}\medskip\noindent{\bf
Remark \theremark:} }{$\Box$\medskip}
\newenvironment{theorem}{\refstepcounter{theorem}
\medskip\noindent{\sc Theorem \thetheorem}:}{$\Box$\medskip}

\newenvironment{lemma}{\refstepcounter{theorem}\medskip\noindent{\sc
Lemma \thetheorem}:}{ $\Box$\medskip }

\newenvironment{definition}{\refstepcounter{definition}\medskip\noindent{\sc
Definition \thedefinition}:}{$\Box$\medskip}

\def\op#1{\mathop{{\it\fam0} #1}\limits}

\newcommand{\wh}{\widehat}

\newcommand{\wt}{\widetilde}
\newcommand{\ol}{\overline}
\newcommand{\ot}{\otimes}

\newcommand{\ar}{\op\longrightarrow}
\newcommand{\id}{{\mathrm{Id}\,}}
\newcommand{\di}{{\mathrm {dim}\,}}
\newcommand{\Ker}{\mathrm{Ker}\,}
\newcommand{\nm}[1]{|{#1}|}
\newcommand{\hm}{{\rm Hom\,}}

\newcommand{\cC}{{\mathcal{C}\ell}}
\newcommand{\cV}{{\mathcal V}}
\newcommand{\cF}{{\mathcal F}}
\newcommand{\cO}{{\mathcal O}}
\newcommand{\cT}{{\mathcal T}}
\newcommand{\cJ}{{\mathcal J}}
\newcommand{\cG}{{\mathcal G}}

\newcommand{\cE}{{\mathcal E}}
\newcommand{\bT}{{\mathbf T}}
\newcommand{\rA}{{\mathrm{Ann}\,}}

\newcommand{\f}{\phi}
\newcommand{\Si}{\Sigma}
\newcommand{\La}{\Lambda}
\newcommand{\si}{\sigma}
\newcommand{\vr}{\varrho}
\newcommand{\al}{\alpha}
\newcommand{\thh}{\theta}
\newcommand{\bt}{\beta}
\newcommand{\g}{\gamma}

\newcommand{\Om}{\Omega}

\newcommand{\m}{\mu}
\newcommand{\dl}{\delta}
\newcommand{\la}{\lambda}
\newcommand{\bb}{{\mathbf 1}}
\newcommand{\dr}{\partial}
\newcommand{\w}{\wedge}
\newcommand{\bL}{{\mathbf L}}

\newcommand{\vt}{\vartheta}
\newcommand{\xx}{\times}
\newcommand{\up}{\upsilon}

\newcommand{\lla}{\op\longleftarrow}

\newcommand{\cA}{{\mathcal A}}
\newcommand{\cK}{{\mathcal K}}
\newcommand{\cZ}{{\mathcal Z}}
\newcommand{\cS}{{\mathcal S}}
\newcommand{\cD}{{\mathcal D}}
\newcommand{\cU}{{\mathcal U}}

\newcommand{\lto}{{\leftarrow}}

\newcommand{\gd}{{\mathfrak d}}

\newcommand{\gf}{{\mathfrak f}}
\newcommand{\e}{\epsilon}

\newcommand{\ve}{\varepsilon}

\newcommand{\mar}[1]{}

\begin{document}

\hbox{}

\begin{center}

{\Large\bf Deformation quantization on jet manifolds}
\bigskip

G.SARDANASHVILY, A.ZAMYATIN

\medskip

Physics Faculty, Moscow State University, Russia

\bigskip

\bigskip

\textbf{Abstract}

\end{center}

Deformation quantization conventionally is described in terms of
multidifferential operators. Jet manifold technique is well-known
provide the adequate formulation of theory of differential
operators. We extended this formulation to the multidifferential
ones, and consider their infinite order jet prolongation. The
infinite order jet manifold is endowed with the canonical flat
connection that provides the covariant formula of a deformation
star-product.

\tableofcontents

\section{Introduction}

Deformation quantization conventionally is described in terms of
multidifferential operators. Jet formalism provides the adequate
formulation of theory of differential operators and differential
equations on manifolds and fibre bundles \cite{bryant,book,kras}.
Therefore, we develop an idea in \cite{cat02,cat04,book05} and aim
to describe deformation quantization in terms of jets.

For this purpose, we develop theory of differential and
multidifferential operators on a ring $C^\infty(X)$ of smooth real
functions on a smooth manifold $X$ as functions on an infinite
order jet manifold $J^\infty F$ (\ref{df105}) (Definition
\ref{df73} and Theorem \ref{df148}). Their Hochschild complex
(\ref{df255}) is constructed, and the deformation (\ref{df181}) is
defined. A key point is that an infinite order jet manifold
$J^\infty F$ is endowed with the canonical flat connection
(\ref{df509}) that provides the covariant formula (\ref{df810}) of
a deformation star-product.

\section{Deformation quantization of Poisson manifolds}

In a general setting, a deformation of an algebra $A$ over a
commutative ring $\cK$ is its Gerstenhaber extension to an algebra
$\cA_h$ (Definition \ref{df4}) over the ring $\cK[[h]]$ of formal
power series in a real number $h>0$, called the deformation
parameter \cite{gerst}. In these terms, deformation quantization
is defined to be a deformation of a Poisson algebra of smooth real
functions on a Poisson manifold where $h$ is treated as a Plank
constant \cite{bayen}.

\subsection{Gerstenhaber's deformation of algebras}

This Section summarizes the relevant material on deformations of
algebraic structures \cite{fial,gerst,gerst92,book05}.

\subsubsection{Formal deformation}

Let $A$ be a (not necessarily associative) algebra. Let us
consider a set $A[[h]]$ of formal power series
\be
a_h=a+h a_1 +h^2 a_2+\cdots, \qquad a_i\in A,
\ee
in a deformation parameter $h$ whose coefficients are elements of
$A$. It naturally is an algebra, called the power series algebra,
with respect to the formal sum and product of power series.

Let $\cK$ be a commutative ring (i.e. a unital associative
algebra), and let $A$ be a $\cK$-algebra. Then $\cK[[h]]$ also is
a commutative ring, and a power series algebra $A[[h]]$ is both a
$\cK[[h]]$-module and a $\cK[[h]]$-algebra. It also is a
$\cK$-algebra, and there is a canonical monomorphism $A\to
A[[h]]$.

A power series algebra $A[[h]]$ can be provided with a different
algebraic structure as follows. Let $\al_r$, $r=1,\ldots$, be a
series of $\cK$-bilinear maps $A\times A\to A$ extended to
$A[[h]]$. Let us define a $\cK[[h]]$ bilinear map multiplication
\mar{053}\beq
a*b=\al_h(a,b)=ab+\op\sum_{r=1} h^r\al_r(a,b), \qquad a,b\in
A[[h]], \label{053}
\eeq
in the $\cK[[h]]$-module $A[[h]]$.

\begin{definition} \label{df4} \mar{df4}
The multiplication (\ref{053}) is called the formal deformation
(or, in short, a deformation) of an original multiplication in
$A$, and makes $A[[h]]$ into a $\cK[[h]]$-algebra $\cA_h$ called
the (formal) deformation of $A$.
\end{definition}

One can think of the multiplication (\ref{053}) as being the power
series
\be
\al_h=\al + \op\sum_{r=1} h^r\al_r,
\ee
of the multiplications $\al_r$ in $A[[h]]$

\begin{remark} \label{df1} \mar{df1}
For instance, the power series algebra $A[[h]]$ with the original
multiplication law $a*b=ab$ is called the null deformation of $A$.
\end{remark}

\begin{remark} \label{df2} \mar{df2}
A $\cK$-algebra $\cA_h$ contains a two-sided ideal generated by
elements of $hA\subset \cA_h$ so that $\cA_h/h\cA_h=A$
\end{remark}

\begin{definition} \label{df231} \mar{df231}
Two deformations $\cA_h$ and $\cA'_h$ of a $\cK$-algebra $A$ are
said to be equivalent if they are isomorphic $\cK[[h]]$-algebras,
i.e., there exists an isomorphism of $\cK[[h]]$-modules
\mar{071}\beq
\f_h:A[[h]]\to A[[h]] \label{071}
\eeq
such that the relation
\mar{070}\beq
\f_h(\al'_h(a,b))=\al_h(\f_h(a),\f_h(b)) \label{070}
\eeq
holds.
\end{definition}

For the sake of brevity, let us write $\al'_h=\al_h\circ \f_h$.
Any $\cK[[h]]$-linear morphism (\ref{071}) is necessarily a formal
power series
\mar{058}\beq
\f_h=\f+h\f_1+h^2\f_2+\cdots, \label{058}
\eeq
whose coefficients are $\cK$-linear maps $\f_i:A\to A$.
Substituting this power series into the relation (\ref{070}), one
easily obtains that
\mar{073}\beq
\f=\id A, \qquad  \f_1(ab)+\al'_1(a,b)=\al_1(a,b)
+\f_1(a)b+a\f_1(b). \label{073}
\eeq

\begin{definition} \label{df232}
A deformation $\cA_h$ of $A$ is said to be trivial if it is
equivalent to the null deformation $A[[h]]$ of $A$.
\end{definition}

\begin{remark} \label{075} \mar{075}
A $\cK[[h]]$-algebra automorphism $\f_h$ of the power series
algebra $A[[h]]$ obeys the relation
\mar{059}\beq
\f_h(ab)=\f_h(a)\f_h(b), \qquad a,b\in A. \label{059}
\eeq
In particular, the second equality (\ref{073}) takes a form
\be
\f_1(ab)=\f_1(a)b+a\f_1(b),
\ee
i.e., $\f_1$ is a derivation of an algebra $A$.
\end{remark}

\begin{definition} \label{df5} \mar{df5}
Let $\cS\to A$ be an algebra morphism. A deformation $\cA_h$ of
$A$ is said to be the $\cS$-relative deformation if
\be
a*s=as, \qquad s*a=sa
\ee
for all $a\in A$ and $s\in \cS$.
\end{definition}

For instance, let $A$ be a unital algebra and $\cS=\{\bb\}$. One
can show that any deformation of $A$ is unital.

\subsubsection{Deformation of rings}

Let $A$ be $\cK$-ring, i.e., a unital associative algebra over a
commutative ring $\cK$. We aim to study its deformations
(\ref{053}) which are rings. It is readily observed that the null
deformation of a ring is a ring. A ring which admits only trivial
deformations is called rigid.

\begin{lemma} \label{df9} \mar{df9}
 One can show the following.

(i) Any deformation of $A$ is equivalent to that where the unit
element coincides with the unit element of $A$.

(ii) Invertible elements of $A$ remain invertible in any
deformation.
\end{lemma}

In order to be associative, a deformation $\cA_h$ of $A$ must
satisfy the associativity condition
\mar{qm885}\ben
&& (a_1*a_2)* a_3- a_1* (a_2* a_3) =
\op\sum_{k=1} h^k D_k(a_1,a_2,a_3)=0, \nonumber\\
&& D_k(a_1,a_2,a_3)= \label{qm885}\\
&& \qquad \op\sum_{s+r=k,\,s,r\geq 0} [\al_r(\al_s(a_1,a_2),a_3) -
\al_r(a_1,\al_s(a_2,a_3))], \nonumber
\een
that is,
\mar{061}\beq
D_k(a_1,a_2,a_3)=0, \qquad k=1,\ldots, \qquad a_i\in A.
\label{061}
\eeq
This condition is phrased in terms of the Hochschild cohomology
(Definition \ref{df16}) as follows.

Let $E_k(a_1,a_2,a_3)$ denote the sum of the terms with indices
$1\leq s,r\leq k$ in the right-hand side of the expression
(\ref{qm885}). It is easily observed that $E_1=0$, while each
$E_k$, $k=2,\ldots,$ depends on the terms $\al_i$,
$i=1,\ldots,k-1$. Then the condition (\ref{061}) takes a form
\mar{qm886}\beq
D_k(a_1,a_2,a_3)=E_k(a_1,a_2,a_3) -(\dl \al_k)(a_1,a_2,a_3)=0,
\label{qm886}
\eeq
where
\mar{qm887}\ben
&& (\dl \al_k)(a_1,a_2,a_3)=a_1\al_k(a_2,a_3)-\al_k(a_1a_2,a_3)+
\label{qm887}\\
&& \qquad \al_k(a_1,a_2a_3)-\al_k(a_1,a_2)a_3 \nonumber
\een
is the Hochschild coboundary operator (\ref{spr106}) of the
Hochschild complex $B^*(A,A)$ (\ref{ws50}). Thus, one can think of
the terms $\al_k$ of the power series (\ref{053}) as being
two-cochains of the above mentioned Hochschild complex $B^*(A,A)$.
Since $E_1=0$, a glance at the condition (\ref{qm886}) shows that
$\al_1$ is a two-cocycle of the Hochschild complex. Then we can
obtain an associative deformation in the framework of the
following recurrence procedure. Let us assume that $\al_i$, $i\leq
k$, are two-cochains such that $D_i=0$ for all $i\leq k$. One can
show that $\dl E_{k+1}=0$, i.e., $E_{k+1}$ is a three-cocycle
whose cohomology class depends only on that of $\al_i$,
$i=1,\ldots,k$. If this cocycle is a coboundary, i.e., it belongs
to the zero element of the Hochschild cohomology group $H^3(A,A)$,
then a desired cochain $\al_{k+1}$ can be found.

For instance, let $\cA_h$ be a deformation of $A$ such that
$\al_i=0$, $i=1,\ldots,k-1$, $k>1$. Then $E_k=0$, and the
condition (\ref{qm886}) shows that $\al_k$ is a two-cocycle. As
was mentioned above, this also is the case of $k=1$. The
Hochschild cohomology class $[\al_k]$ of $\al_k$ (or, if there is
no danger of confusion, $\al_k$ itself) is called the
infinitesimal of a deformation.  Let us assume that $\al_k=\dl \f$
is a coboundary. It is easily verified that
\mar{076}\beq
\al'_h=\al_h\circ (\id A-h^k\f) \label{076}
\eeq
is an equivalent deformation such that $\al'_i=0$, $i=1,\ldots,k$.
If $H^2(A,A)=0$ and $\al_h$ is a deformation, one can use the
equivalences (\ref{076}) in order to remove all the terms $\al_i$
as follows.

\begin{theorem} \label{093} \mar{093}
Equivalent deformations of $A$ possess the same infinitesimal
$[\al_1]\in H^2(A,A)$.
 \end{theorem}

 \begin{theorem} \label{077} \mar{077}
If $A$ is a $\cK$-ring with $H^2(A,A)=0$, then any deformation of
$A$ is trivial, i.e., $A$ is rigid.
 \end{theorem}

A ring $A$ is called absolutely rigid if $H^2(A,A)=0$.

Let $H^3(A,A)=0$. For any two-cocycle $\al$, there exists a
deformation of $A$ such that $\al_1=\al$. Indeed, $\al_1=\al$
defines $E_2$ by the formulae (\ref{qm885}) and (\ref{qm886}).
This three-cocycle is a coboundary. Therefore, there exists a
two-cochain $\al_2$ such that the term $D_2$ (\ref{qm886})
vanishes. The two-cochains $\al_1$ and $\al_2$ define the
three-cocycle $E_3$ by the formulae (\ref{qm885}) and
(\ref{qm886}). It also is a coboundary. Consequently, there exists
a two-cochain $\al_3$ such that the term $D_3$ (\ref{qm886})
vanishes, and so on. Thus, elements of the Hochschild cohomology
group $H^3(A,A)$ provide the obstruction to a Hochschild
two-cocycle be the infinitesimal of a deformation.

\begin{example} \label{0120} \mar{0120}
Let $\cK$ be a $\mathbb Q$-ring.  Let $u$ and $v$ be derivations
of a $\cK$-ring $A$. They are Hochschild one-cocycles. Their
cup-product $u\smile v$ (\ref{ws71}) is a two-cocycle. This
two-cocycle need not be the infinitesimal of a deformation of $A$,
unless $u$ and $v$ mutually commute. If the derivations $u$ and
$v$ commute, they define a deformation of $A$ given by the formula
\be
a*b=\exp(hu\smile v)(a,b) = ab + \op\sum_{r=1}
\frac{h^r}{r!}u^r(a)v^r(b), \qquad a,b\in A.
\ee
\end{example}

\subsection{Star-product}

A Poisson bracket on the ring $C^\infty(Z)$ of smooth real
functions on a manifold $Z$ (or a Poisson structure on $Z$) is
defined as an $\mathbb R$-bilinear map
\be
C^\infty(Z)\times C^\infty(Z)\ni (f,g)\to \{f,g\}\in C^\infty(Z)
\ee
which satisfies the following conditions:

$\bullet$ $\{g,f\}=-\{f,g\}$ (skew-symmetry);

$\bullet$ $\{f,\{g,h\}\} + \{g,\{h,f\}\} +\{h,\{f,g\}\}=0$ (the
Jacobi identity);

$\bullet$ $\{h,fg\}=\{h,f\}g +f\{h,g\}$ (the  Leibniz rule).

A manifold $Z$ endowed with a Poisson structure is called a
Poisson manifold.  A Poisson bracket makes $C^\infty(Z)$ into a
real Lie algebra, called the Poisson algebra. A Poisson structure
is characterized by a particular bivector field as follows.

\begin{theorem}\label{p11.1} \mar{p11.1} Every Poisson bracket on a
manifold $Z$ is uniquely defined as
\mar{b450}\beq
\{f,f'\}=w(df,df')=w^{\m\nu}\dr_\m f\dr_\nu f' \label{b450}
\eeq
by the bivector field $w$ (\ref{df92}) whose Schouten--Nijenhuis
bracket $[w,w]_\mathrm{SN}$ (\ref{df94}) vanishes. It is called a
Poisson bivector field \cite{vais}.
\end{theorem}

Let $(Z,\{,\})$ be a Poisson manifold. By a star-product
 on $Z$ is meant a ring deformation
\mar{0115}\beq
f* f'=ff'+ \op\sum_{r=1} h^r\al_r(f,f') \label{0115}
\eeq
of a real ing $C^\infty(Z)$ of smooth real functions on $Z$ such
that
\mar{0130}\beq
\al_1(f,f')-\al_1(f',f)=2\{f,f'\}, \qquad f,f'\in C^\infty(Z).
\label{0130}
\eeq
In accordance with Lemma \ref{df9}, one can always choose $*$ such
that
\be
f*1=1*f=f, \qquad f\in C^\infty(Z).
\ee

\begin{remark} \label{df300} \mar{df300}
Given the star-product (\ref{0115}) -- (\ref{0130}), the
commutator
\mar{0117}\ben
&& [f,f']_h=(2h)^{-1}(f* f'-f'* f)=\{f,f'\}+ \label{0117}\\
&& \qquad \frac12\op\sum_{r=1} h^r(\al_{r+1}(f,f')-
\al_{r+1}(f',f)) \nonumber
\een
provides a Lie deformation of the Poisson bracket $\{,\}$ on $Z$.
This deformation is treated as deformation quantization.
\end{remark}

\begin{remark}
Here, we are not concerned with the Laurent series
$\cK[h^{-1},h]]$ in $h$, i.e., a polynomial in $h^{-1}$ and a
formal series in $h$.
\end{remark}

Hereafter, we restrict our consideration to differential
deformations (\ref{0115}) where $\al_r(f,f')$ are bilinear
differential (bidifferential) operators of finite order. They are
equivalent to continuous deformations as follows (Theorem
\ref{0129}).

\begin{remark} \label{ws2} \mar{ws2}
Whenever referring to a topology on the ring $C^\infty(X)$, we
will mean the topology of compact convergence for all derivatives
\cite{rob}. The $C^\infty(X)$ is a Fr\'echet ring with respect to
this topology, i.e., a complete  metrizable locally convex
topological vector space.
\end{remark}

Because $C^\infty(Z)$ is a Fr\'echet ring (Remark \ref{ws2}), by
its  continuous deformation is meant the deformation (\ref{0115})
where bilinear maps $\al_r$ are continuous in each argument.
Continuous cochains form a subcomplex
$B^*_\mathrm{c}(C^\infty(Z),C^\infty(Z))$ of the Hochschild
complex $B^*(C^\infty(Z),C^\infty(Z))$ (\ref{ws50}). Let
$H^*_\mathrm{c}(C^\infty(Z),C^\infty(Z))$ denote its cohomology,
called the continuous Hochschild cohomology. Another subcomplex
$B^*_\mathrm{d}(C^\infty(Z),C^\infty(Z))$ of the Hochschild
complex $B^*(C^\infty(Z),C^\infty(Z))$ consists of cochains which
are $C^\infty(X)$-valued multidifferential operators on
$C^\infty(X)$ of finite order. Let us denote
\mar{df288}\beq
\cD^*=B^*(C^\infty(Z),C^\infty(Z)). \label{df288}
\eeq
Since every differential operator is continuous with respect to
the Fr\'echet topology of $C^\infty(Z)$, a complex $\cD^*$ is a
subcomplex of a complex $B^*_\mathrm{c}(C^\infty(Z),C^\infty(Z))$
of continuous Hochschild cochains. Cohomology $H^*(\cD^*)$ of the
complex $\cD^*$ is called the differential Hochschild cohomology.
One can show that continuous and differential Hochschild
cohomology groups are isomorphic \cite{nad99,pelaum,pinc1}:
\be
H^r_\mathrm{c}(C^\infty(Z),C^\infty(Z))=H^r(\cD^*).
\ee
As a consequence, one comes to the following result
\cite{nad99,pinc1}.

\begin{theorem} \label{0129} \mar{0129}
Every continuous deformation of $C^\infty(Z)$ is equivalent to the
differential one. A continuous equivalence between two
differential deformations of $C^\infty(Z)$ is differential (i.e.,
the maps $\f_i$ (\ref{058}) are linear differential operators).
\end{theorem}

Let us turn to the star-product $*$ (\ref{0115}) -- (\ref{0130})
on a Poisson manifold $(Z,\{,\})$ defined by the differential
deformation (\ref{0115}) of a real ring $C^\infty(Z)$. It is a Lie
deformation of the Poisson algebra $C^\infty(Z)$. Of course, an
arbitrary deformation (\ref{0115}) need not be a star-product. A
key point is that, in contradistinction with the Hochschild
cohomology, the needed Chevalley--Eilenberg cohomology is small
\cite{vey}.

\begin{lemma} \label{0122} \mar{0122}
Any multivector field $\vt\in \cT_r(Z)$ (\ref{cc6}) of degree $r$
defines a differential Hochschild $r$-cocycle
\mar{0123}\beq
\vt(f_1,\ldots,f_r) =\vt\lfloor df_1\w\cdots\w df_r\in \cD^r,
\label{0123}
\eeq
where $\lfloor$ is the right interior product (\ref{df11}).
\end{lemma}

Thus, there is the inclusion
\mar{0124}\beq
\cT_r(Z)\subset \cD^*. \label{0124}
\eeq
Moreover, any differential Hochschild cocycle is cohomologous to
some multivector field (\ref{0123}) as follows \cite{hoch,vey}.

\begin{theorem} \label{0125} \mar{0125}
The inclusion (\ref{0124}) induces  $C^\infty(Z)$-module
isomorphisms
\mar{0126}\beq
H^r(\cD^*)=\cT_r(Z). \label{0126}
\eeq
\end{theorem}

\begin{remark} \label{0150} \mar{0150}
By virtue of Theorem \ref{0125}, any Hochschild two-cocycle is
cohomologous to its skew-symmetric part. Therefore, we can
restrict our consideration to deformations of $C^\infty(Z)$ whose
infinitesimal is the Poisson bracket
\mar{0151}\beq
\al_1(f,f')=\{f,f'\}, \qquad f,f'\in C^\infty(Z), \label{0151}
\eeq
in order to obtain a star-product. The fact that a regular Poisson
bracket is a Hochschild two-cocycle is easily justified when it is
written with respect to local Darboux coordinates $(p_i,q^i)$ and,
thus, is a sum of cup-products $\dr^i\smile\dr_i$ of mutually
commutative vector fields $\dr^i$ and $\dr_i$ (Example
\ref{0120}).
\end{remark}

Let us consider star-products on a symplectic manifold $(Z,\Om)$.

The Moyal product on $Z=\mathbb R^{2m}$ was the first example of a
star-product \cite{han,kamm}.  Let $\mathbb R^{2m}$ be provided
with the coordinates $(q^i,p_i)$ and the canonical symplectic form
$\Om=dp_i\w dq^i$. Let us consider a differential ring deformation
of $C^\infty(\mathbb R^{2m})$ whose infinitesimal is the Poisson
bracket (\ref{0150}). This infinitesimal is a sum of cup-products
of mutually commutative vector fields $\dr^i$ and $\dr_i$. Then,
generalizing  Example \ref{0120}, one can show that such a
deformation exists and it is given by  the expression
\mar{qm893}\ben
&& f* f'=\exp[\frac{h}{2}\{f,f'\}]=
f\exp[\frac{h}{2}(\op\dr^{\lto i}{\op\dr^\to}_i-
{\op\dr^\lto}_i\op\dr^{\to i})]f'=
\label{qm893}\\
&& \qquad \op\sum_{k=0}\frac{h^k}{k!} \op\sum_{r=0}^k
(-1)^r(\dr_{i_1}\cdots \dr_{i_r}\dr^{i_{r+1}}\cdots\dr^{i_k}f)
(\dr^{i_1}\cdots \dr^{i_r}\dr_{i_{r+1}}\cdots\dr_{i_k}f').
\nonumber
\een
This is a star-product, called the Moyal product.
 Since the de Rham cohomology group
$H^2_\mathrm{DR}(\mathbb R^{2m})$ of $\mathbb R^{2m}$ is trivial,
all star-products on $(\mathbb R^{2m}, \Om)$ are equivalent to the
Moyal one (\ref{qm893}). This star-product defines the
corresponding Lie deformation (\ref{0117}).

Fedosov's deformation quantization \cite{fed} generalizes a
construction of the Moyal product (\ref{qm893}) to an arbitrary
symplectic manifold as follows.

\begin{theorem} \label{0131} \mar{0131}
Any symplectic manifold admits a star-product \cite{wild}, e.g.,
Fedosov's one \cite{fed}.
\end{theorem}

\begin{theorem} \label{0132} \mar{0132}
Any star-product on a symplectic manifold is equivalent to some
Fedosov star-product \cite{bert97,xu98}.
\end{theorem}

\begin{theorem} \label{0133} \mar{0133}
The equivalence classes of star-products on a symplectic manifold
constitute an affine space modelled on the linear space
$H^2(Z)[[h]]$ of power series in $h$ whose coefficients are
elements of the de Rham cohomology group $H^2_\mathrm{DR}(Z)$ of
$Z$ \cite{delign,gutt,nest}.
\end{theorem}

Namely, given two different star-products $*$ and $*'$, one
associate to them a unique \v Cech cohomology class
\mar{0134}\beq
t(*',*)\in H^2(Z;\mathbb R)[[h]], \label{0134}
\eeq
called Deligne's relative class. It is defined as follows
\cite{delign,gutt}.

\begin{lemma} \label{0135} \mar{0135}
Star-products $*$ and $*'$ are equivalent if they are equivalent
as ring deformations, i.e.,
\mar{0137}\ben
&& \f(f*g)=\f f*'\f g, \qquad f,g\in C^\infty(Z), \nonumber\\
&& \f=\id +\op\sum_{r=1} h^r\f_r, \label{0137}
\een
where $\f_r$ are differential operators.
\end{lemma}

One calls $\f$ (\ref{0137}) the formal differential operator.

\begin{theorem} \label{0138} \mar{0138}
Let $(Z,\Om)$ be a symplectic manifold, and let us suppose that
the \v Cech cohomology group $H^2(Z;\mathbb R)$ is trivial. Then
any two star-products on $Z$ are equivalent.
\end{theorem}

\begin{theorem} \label{0139} \mar{0139}
Let $*$ be a star-product on $(Z,\Om)$, and let us suppose that
$H^1(Z;\mathbb R)=0$. Then any self-equivalence $\f$ (\ref{0137})
of $*$ is inner, i.e.,
\be
\f=\exp\{\mathrm{ad}_* \g\},
\ee
for some $\g\in C^\infty(Z)[[h]]$, where
\be
\mathrm{ad}_*\g(f) =[g,f]_*=g*f-f*g, \qquad f\in C^\infty(Z)[[h]],
\ee
is the star adjoint representation.
\end{theorem}

 Let $\{U_i\}$ be a locally finite open cover of $Z$ by
Darboux coordinate charts such that $U_i$ and all their non-empty
intersections are contractible. Let us denote $C_i=C^\infty(U_i)$.
A star-product $*$ in $C^\infty(Z)[[h]]$ certainly yields a
star-product $*$ in $C_i$ on a symplectic manifold $(U_i,\Om)$.
Let $*$ and $*'$ be two star-products on $Z$. By virtue of Theorem
\ref{0138}, their restrictions to $U_i$ are equivalent star
products, i.e., there exists the formal differential operator
(\ref{0137}):
\be
\f_i:C_i[[h]]\to C_i[[h]]
\ee
such that
\be
\f_i(f*g)=\f_if*\f_ig, \qquad f,g\in C_i[[h]].
\ee
On $U_i\cap U_j$, we accordingly have a self-equivalence
$\f_j^{-1}\circ \f_i$ of $*$ in $C_{ij}[[h]]$. By virtue of
Theorem \ref{0139}, this self-equivalence is inner, i.e., there
exists an element $\g_{ji}=-\g_{ij}\in C_{ij}[[h]]$ such that
\be
\f_j^{-1}\circ \f_i=\exp\{\mathrm{ad}_* \g_{ji}\}.
\ee
On $U_i\cap U_j\cap U_k$, the composition
\be
\mathrm{ad}_*\g_{kji}=\mathrm{ad}_*\g_{ik}\circ\mathrm{ad}_*\g_{kj}\circ
\mathrm{ad}_*\g_{ji}
\ee
is the identity morphism of $C_{ijk}[[h]]$ and, consequently, is
represented by an element $\g_{kji}$ in the center $\mathbb
R[[h]]$ of $C_{ijk}[[h]]$. The standard arguments show that the
set of the elements $\g_{kji}$ define a \v Cech two-cocycle  with
values in $\mathbb R[[h]]$. Its cohomology class $[\g_{kji}]\in
H^2(Z;\mathbb R)$ is desired Deligne's relative class
(\ref{0134}).

\begin{theorem} \label{0143} \mar{0143}
If $*$, $*'$ and $*''$ are three star-products on $(Z,\Om)$, then
\be
t(*'',*)=t(*'',*') +t(*',*).
\ee
\end{theorem}

A moduli space of equivalent star-products on a symplectic
manifold $(Z,\Om)$ is usually identified with
\be
\frac{1}{h}[\Om] +H^2(Z)[[h]].
\ee
Let us point out an isomorphism
\mar{0141}\beq
H^2(C^\infty(Z)[[h]],C^\infty(Z)[[h]])=Z^2(Z) + H^2(Z)[[h]],
\label{0141}
\eeq
where $Z^2(Z)$ is a space of closed two-forms on $Z$.

\subsection{Kontsevich's deformation quantization}

Let us turn to star-products on Poisson manifolds. A star-product
on an arbitrary (smooth, not necessarily regular) Poisson manifold
exists. Its general construction was first suggested in
\cite{konts0,konts2}.

If $(Z,\{,\}$ is a regular Poisson manifold, there exists a
tangential star-product on $Z$ \cite{fed1,masm}. It is defined as
follows. Let $\cS_\cF(Z)$ be the center of the Poisson algebra
$C^\infty(Z)$. Let us consider a $\cS_\cF(Z)$-relative deformation
of a real ring $C^\infty(Z)$. A star-product on $Z$ and a Lie
deformation of $C^\infty(Z)$) which comes from an
$\cS_\cF(Z)$-relative deformation (Definition \ref{df5}) are
called tangential because $\cS_\cF(Z)$ consists of functions
constant on leaves of the characteristic foliation of $(Z,\{,\})$.
A tangential star-product can be introduced either as the
tangential version of Vey's work \cite{lich82,masm} or the
straightforward generalization of Fedosov's deformation
quantization of symplectic manifolds to symplectic foliations.

Kontsevich's deformation \cite{konts0,konts2} generalizes the
Moyal star-product on $\mathbb R^{2m}$ to a generic Poisson
structure on $\mathbb R^{2m}$ which fails to be regular and does
not admit the Darboux coordinates. Recently, Kontsevich's
deformation quantization has been extended to an arbitrary Poisson
manifold \cite{cat02,dolg,konts}. The key point of Kontsevich's
deformation is the formality theorem (Theorem \ref{0221}). This
theorem and Theorem \ref{0240} establish the relations between
algebras of multivector fields and multidifferential operators on
a smooth manifold.

\subsubsection{Differential graded Lie algebras}

We start with the relevant algebraic constructions (see, e.g.,
\cite{cat04,dito}). Unless otherwise stated, all algebras are over
a field $\mathbb K$ of characteristic zero. By a graded structure
is meant a $\mathbb Z$-graded structure, and the symbol $|.|$
stands for the $\mathbb Z$-graded parity.

A  differential graded Lie algebra  (henceforth a DGLA)
 is a differential graded algebra
\be
\cG=\op\oplus_{i\in \mathbb Z} \cG^i,
\ee
whose multiplication operation is  a  graded Lie bracket
\be
[,]:\cG\ot\cG\to\cG, \qquad [a,b]\in \cG^{|a|+|b|},
\ee
satisfying the relations
\mar{0211,2}\ben
&& [a,b]=-(-1)^{|a||b|}[b,a], \qquad a,b,c\in\cG, \nonumber\\
&& (-1)^{|a||c|}[a,[b,c]] + (-1)^{|b||a|}[b,[c,a]]
+(-1)^{|c||b|}[c,[a,b]]=0, \label{0211}
\een
and the differential $d$ of degree one obeys the graded  Leibniz
rule
\mar{0212}\beq
d[a,b]=[da,b]+(-1)^{|a|}[a,db]. \label{0212}
\eeq
The degree zero part $\cG^0$ and the even part of a DGLA $\cG$ are
Lie algebras. Any Lie algebra is a DGLA of zero degree with $d=0$.

A morphism of DGLAs  is a graded linear map of degree zero which
commutes with differentials and brackets. In particular, it is a
cochain morphism.

Let $H^*(\cG)$ be the cohomology  of a DGLA $\cG$ (Definition
\ref{df15}).

\begin{theorem} \label{df272} \mar{df272}
The cohomology $H^*(\cG)$   of a DGLA $\cG$ is a DGLA with respect
to the zero differential $d=0$ and the bracket
\be
[\ol a,\ol b]_H =\ol{[a,b]}, \qquad a,b\in\cG,
\ee
where $\ol a$ denotes the cohomology class of $a\in\cG$.
\end{theorem}

It is evident that the cohomology  of a DGLA $H^*(\cG)$ coincides
with $H^*(\cG)$, i.e., $H^*(H^*(\cG))=H^*(\cG)$.

Every morphism $\Phi:\cG\to\cG'$ of DGLAs yields a morphism of the
cohomology DGLAs
\mar{0191}\beq
\ol \Phi: H^*(\cG)\to H^*(\cG'). \label{0191}
\eeq

\begin{definition} \label{df275} \mar{df275}
A DGLA morphism $\Phi$ is called the quasi-isomorphism if the
induced morphism $\ol\Phi$ (\ref{0191}) is an isomorphism.
\end{definition}

Let us emphasize that the existence of a quasi-isomorphism
$\cG\to\cG'$ does not imply the existence of the 'quasi-inverse'
$\cG'\to\cG$ which induces the inverse morphism $\ol\Phi^{-1}$.

\begin{definition} \label{df276} \mar{df276}
If the quasi-inverse exists, the DGLAs $\cG$ and $\cG'$ are called
quasi-isomorphic. A DGLA $\cG$ is called formal if it is
quasi-isomorphic to the DGLA $H^*(\cG)$.
\end{definition}

A graded coalgebra is a graded vector space $\gf$ provided with
the following graded co-operations:

$\bullet$ a graded comultiplication $\Delta: \gf\to \gf\ot\gf$
such that
\be
\Delta(\gf^i)\subset \op\oplus_{j+k=i}\gf^j\ot\gf^k;
\ee

$\bullet$ a counit $\e:\gf\to\mathbb K$ such that
$\e(\gf^{i>0})=0$.

\noindent These operations obey the relations
\be
(\Delta\ot\id)\circ\Delta=(\id\ot\Delta)\circ\Delta, \quad
(\e\ot\id)\circ\Delta=(\id\ot\e)\circ\Delta=\id.
\ee

A graded coalgebra is called graded cocommutative
 if $P\circ\Delta=\Delta$,
where
\be
P(a\ot b)=(-1)^{|a||b|}b\ot a, \qquad a,b\in\gf,
\ee
is the graded transposition operator.

A coderivation of degree $r$ on a graded coalgebra $\gf$ is a
graded linear map $\dr:\gf^i \to\gf^{i+r}$ which satisfies the
co-Leibniz rule:
\be
(\Delta\circ \dr)(v)=((\dr\ot\id +(-1)^{r|v|} \id\ot\dr)\circ
\Delta) (v), \qquad v \in\gf.
\ee
By a  codifferential on a graded coalgebra is meant a nilpotent
coderivation of degree one.

\subsubsection{$L_\infty$-algebras}

The fact that the quasi-inverse of DGLAs need not exists has
motivated one to call into play a wider class of algebras, which
are $L_\infty$-algebras.

Let $V$ be a graded vector space and
\be
\ot V=\mathbb K\oplus V\oplus\cdots\oplus V^{\ot k}\oplus\cdots
\ee
its tensor algebra.  It is a graded vector space such that
\be
|v\ot v'|=|v|+|v'|, \qquad v,v'\in V,
\ee
and $|\la|=0$, $\la\in\mathbb K$. We call $\ot_+ V= \ot V\setminus
\mathbb K$ the reduced tensor algebra.

A tensor algebra $\ot V$ is brought into a graded coalgebra with
respect to the counit $\e:\ot V\to\mathbb K$ and the
comultiplication
\be
&& \Delta(v_1\ot\cdots \ot v_r)=1\ot(v_1\ot\cdots \ot v_r)+\\
&& \qquad \op\sum^{r-1}_{j=1} ((v_1\ot\cdots \ot
v_j)\ot((v_{j+1}\ot\cdots \ot v_r) +(v_1\ot\cdots \ot v_r)\ot 1.
\ee
A graded symmetric algebra $\vee V$ and a graded exterior algebra
$\w V$ are defined as the the quotients of $\ot V$ by the
two-sided ideals generated by homogeneous elements of the form
$v\ot v'-P(v\ot v')$ and $v\ot v'+P(v\ot v')$, respectively.
Accordingly, we have the reduced algebras $\vee_+V$ and $\w_+V$. A
graded symmetric algebra $\vee V$ is brought into a graded
coalgebra provided with the comultiplication given on elements of
$V$ by
\be
\Delta(v)=1\ot v+v\ot 1,
\ee
and extended as an algebra homomorphism with respect to the tensor
product.

Given a graded  vector space $V$, one can obtain a new graded
vector space $V[k]$ by shifting the degree by $k$, i.e.,
\be
V[k]^i=V^{i+k}, \qquad |v|[k]=|v|-k, \qquad v\in V.
\ee
Then we have the d\'ecalage isomorphism  between the graded
symmetric and exterior algebras. It is given on the $k$-symmetric
power of $V$ shifted by one by the expression
\mar{0192}\ben
&& \zeta_k: \op\vee^kV[1]\to \op\w^kV[k], \label{0192}\\
&& \zeta_k: v_1\vee\cdots\vee v_k\to
(-1)^{\sum^k_{i=1}(k-i)(|v_i|-1}v_1\w\cdots\w v_k. \nonumber
\een

An  $L_\infty$-algebra is a graded vector space $\cG$ endowed with
a codifferential $D$ on the reduced symmetric space
$\vee_+\cG[1]$.

Let us abbreviate  the symbol $D^i_j$ with the projection of $D$
to the component ${\op\vee^i}_+\cG[1]$ of the target vector space
and ${\op\vee^j}_+\cG[1]$ of the domain space. A coderivation $D$
is proved to be uniquely determined by its components $D^1_r$, and
it is a codifferential iff
\mar{0195}\beq
\op\sum_{i=1}^n D^1_i\circ D^i_n=0, \qquad n\geq 1. \label{0195}
\eeq
In particular, $D^1_1\circ D_1^1=0$, i.e., a codifferential $D$
defines a complex on $\cG$.

One can show that any DGLA $(\cG,d)$ can be brought into an
$L_\infty$-algebra where
\be
D^1_1a =(-1)^{|a|}da, \quad D^1_2(a\vee b) =
(-1)^{|a|(|b|-1)}[a,b],\quad D^1_n=0, \quad n\geq 3.
\ee

An $L_\infty$-algebra morphism $\Phi: (\cG,D)\to (\cG',D')$ is a
morphism of graded coalgebras
\mar{0196}\beq
 \Phi: \vee_+\cG[1]\to
\vee_+\cG'[1], \label{0196}
\eeq
which intertwines the codifferentials, i.e., $\Phi\circ
D=D'\circ\Phi$. An $L_\infty$-algebra morphism is uniquely
determined by its components
\mar{0197}\beq
\Phi^1_j: {\op\vee^j}_+\cG[1]\to \cG'[1] \label{0197}
\eeq
which obey the relations
\mar{0198}\beq
\op\sum_{i=1}^n \Phi^1_i\circ D^i_n= \op\sum_{i=1}^n
D'^1_i\circ\Phi^i_n, \qquad n\geq 1. \label{0198}
\eeq

In particular, a morphism $\Phi$ of DGLAs induces their morphism
$\wt\Phi$ as $L_\infty$-algebras such that $\wt \Phi^1_1=\Phi$.

Due to the d\'ecalage isomorphism (\ref{0192}), the codifferential
$D$ defines a sequence $\{l_k\}$ of morphisms
\be
l_k:\op\w^k\cG\to\cG[2-k], \qquad k\geq 1.
\ee
The relation (\ref{0195}) puts an infinite family of conditions on
the $l_k$'s. These conditions imply that $l_1:\cG\to \cG[1]$
 is a differential of degree one obeying the graded Leibniz rule.
 It follows that $(\cG,l_1)$ is a
 complex. The morphism $l_2$ defines a graded bracket
$[.,.]:\op\w^2\cG\to \cG$  on $\cG$ compatible with $l_1$ (i.e.,
the relation (\ref{0212}) holds).
 This bracket obeys the graded Jacobi identity up to homotopy
 given by $l_3$.
 Hence,
 any $L_\infty$-algebra $(\cG,D)$ such that $l_k=0$ for $k\geq
 3$ is a DGLA.

Every $L_\infty$-algebra morphism $\Phi$ provides a cochain
morphism $\Phi_1$ between the complexes $(\cG,l_1)$ and
$(\cG',l'_1)$. One says that an $L_\infty$-morphism $\Phi$
(\ref{0196}) of $L_\infty$-algebras is a quasi-isomorphism if
$\Phi_1$ is a quasi-isomorphism of complexes $(\cG,l_1)$ and
$(\cG',l'_1)$, i.e., it yields an isomorphism of their cohomology.
In contrast with morphisms of DGLAs, any quasi-isomorphism of
$L_\infty$-algebras possesses the quasi-inverse. Thus,
quasi-isomorphisms of $L_\infty$-algebras define an equivalence
relation, i.e., two $L_\infty$-algebras are quasi-isomorphic iff
there is an $L_\infty$-quasi-isomorphism between them. The notion
of formality of $L_\infty$-algebras is formulated similarly to
that of DGLAs.

\subsubsection{Formality theorem}

Let us show that multivector fields on a smooth manifold $Z$
constitute a DGLA. The graded commutative algebra $\cT_*(Z)$ of
multivector fields is customarily provided with the
Schouten--Nijenhuis bracket $[.,.]_\mathrm{SN}$ (\ref{z30}) which
obey the relations (\ref{gm6}) -- (\ref{gm8}). However, there is
another sign convention used in the definition of the
Schouten--Nijenhuis bracket \cite{marl}. This bracket, denoted by
$[.,.]_\mathrm{SN'}$, is
\mar{gm5}\beq
[\vt,\up]_\mathrm{SN'}=-(-1)^{\nm\vt}[\vt,\up]_\mathrm{SN}.
\label{gm5}
\eeq
The relation (\ref{gm6}) for this bracket reads
\mar{gm6'}\beq
[\vt,\up]_\mathrm{SN'}=-(-1)^{(\nm\vt-1)(\nm\up-1)}[\up,\vt]_\mathrm{SN'}.
\label{gm6'}
\eeq
The relation (\ref{gm7}) keeps its form, while the relation
(\ref{gm8}) is replaced with the following one
\mar{gm8'}\ben
&& (-1)^{(\nm\nu-1)(\nm\up-1)}[\nu,[\vt,\up]_\mathrm{SN'}]_\mathrm{SN'}
+ (-1)^{(\nm\vt-1)(\nm\nu-1)}[\vt,[\up,\nu]_\mathrm{SN'}]_\mathrm{SN'}  \nonumber \\
&& \qquad +(-1)^{(\nm\up-1)(\nm\vt-1)}[\up,[\nu,\vt]_\mathrm{SN'}]_\mathrm{SN'}=0. \label{gm8'}
\een
The equalities (\ref{gm6'}) and (\ref{gm8'}) show that, with the
modified Schouten--Nijenhuis bracket (\ref{gm5}), the graded
vector space
\mar{0220}\beq
\cV^*=\cT_*(Z)[1] \label{0220}
\eeq
 of multivector fields on a manifold $Z$ is
precisely a graded Lie algebra. This graded Lie algebra is brought
into a DGLA by setting the differential $d$ to be identically
zero. Clearly, this DGLA is formal.

In particular, let $(Z,w)$ be a Poisson manifold. Then the Poisson
bivector $w$ obeys the Maurer--Cartan equation
\mar{0205}\beq
dw +\frac12[w,w]_\mathrm{SN'}=0 \label{0205}
\eeq
on the DGLA $\cV^*$ (\ref{0220}).

Let us describe the DGLA of multidifferential operators on a ring
$C^\infty(Z)$ of smooth real functions on a manifold $Z$.

In a general setting, let $\cA$ be a $\mathbb K$-ring and
$B^*(\cA,\cA)$ its Hochschild complex (\ref{ws50}). Let us
consider the complex
\mar{0206}\beq
\cC^*= B^*(\cA,\cA)[1]. \label{0206}
\eeq
It inherits the Hochschild coboundary operator (\ref{spr103}):
\mar{0207}\ben
&&(\dl\f^k)(a_0,\ldots,a_{k+1})
= a_0\f^k(a_1,\ldots,a_{k+1}) + \label{0207}\\
&& \quad  \op\sum_j(-1)^{j+1}\f^k
(a_0,\ldots,a_ja_{j+1},\ldots,a_{k+1}) +
(-1)^{k+2}\f^k(a_0,\ldots,a_k)a_{k+1}, \nonumber
\een
the composition product (\ref{ws70}):
\mar{0208}\ben
&& \f^m\circ \f^n(a_0,\ldots,a_{m+n})= \label{0208}\\
&&\quad \op\sum_{i=0}^m (-1)^{in}
\f^m(a_0,\ldots,a_{i-1},\f^n(a_i,\ldots, a_{n+i}),
a_{n+i+1},\ldots, a_{m+n}), \nonumber
\een
and the Gerstenhaber bracket (\ref{ws73}):
\mar{0209}\beq
[\f,\f']_\mathrm{G}=\f\circ \f' -(-1)^{|\f||\f'|}\f'\circ \f,
\qquad \f,\f'\in \cC^*. \label{0209}
\eeq
One can show that this bracket obeys the graded Jacobi identity
(\ref{0211}). Thus, $(\cC^*,[.,.]_\mathrm{G})$ is a graded Lie
algebra. Furthermore, let a one-cocycle
\be
\f^1(a_0,a_1)={\mathfrak m}(a_0,a_1)=a_0a_1
\ee
be the multiplication in $\cA$. We consider the operator
\mar{0210}\beq
d_{\mathfrak m} \f=[{\mathfrak m},\f]_\mathrm{G}, \qquad
\f\in\cC^*, \label{0210}
\eeq
of degree one on $\cC^*$. A direct computation shows that
\be
d_{\mathfrak m}\f=(-1)^{|\f|}\dl\f, \qquad \f\in \cC^*,
\ee
and the relation (\ref{0212}) holds. Then the complex $\cC^*$
(\ref{0206}) is a DGLA with respect to the bracket (\ref{0209})
and the differential (\ref{0210}).

Let $\cA=C^\infty(Z)$, and let us consider a subcomplex $\cD^*$
(\ref{df288}) of the complex $B^*(C^\infty(Z),C^\infty(Z))[1]$
whose cochains are multidifferential operators on $C^\infty(Z)$.
This subcomplex is closed with respect both to the Gerstenhaber
bracket (\ref{0209}) and the action of $d_{\mathfrak m}$. Thus, it
is a desired DGLA of multidifferential operators.

Given a bidifferential operator $\al\in \cD^1$, one can think of
\mar{0213}\beq
f*f'=({\mathfrak m}+\al)(f,f')=ff'+\al(f,f'), \qquad f,f'\in
C^\infty(Z), \label{0213}
\eeq
as being a deformation of the original product $\mathfrak m$ in
$C^\infty(Z)$. One can show that the associativity constraint on
this deformation is given by the equality
\be
[{\mathfrak m}+\al,{\mathfrak m}+\al]_\mathrm{G}=0,
\ee
which takes a form of the Maurer--Cartan equation
\mar{0214}\beq
d_{\mathfrak m}\al+\frac12[\al,\al]_\mathrm{G}=0. \label{0214}
\eeq

Now, the goal is to construct a morphism of the DGLA $\cV^*$
(\ref{0220}) of multivector fields to the DGLA $\cD^*$ of
multidifferential operators which intertwines their differential
graded Lie algebra structures and solutions of the Maurer--Cartan
equations (\ref{0205}) and (\ref{0214}).

One has proved the following \cite{hoch}.

\begin{theorem} \label{0240} \mar{0240}
For any smooth manifold $Z$, there is an isomorphism between the
cohomology $H^*(\cD^*)$ of the algebra $\cD$ and the algebra
$\cV^*$. Since $\cV^*$ coincides with its cohomology $H^*(\cV^*)$,
we have an isomorphism
\mar{0221}\beq
H^*(\cD^*)\to \cV^*\cong H^*(\cV^*). \label{0221}
\eeq
\end{theorem}

The next step is the above mentioned formality theorem.

\begin{theorem} \label{0222} \mar{0222}
The DGLA $\cD^*$ of multidifferential operators on a smooth
manifold $Z$ is formal.
\end{theorem}

It follows that there exists a quasi-isomorphism of the DGLA
$\cD^*$ to $H^*(\cD^*)\cong \cV^*$ and, consequently, there is the
inverse $L_\infty$-quasi-isomorphism $\cU$ of $\cV^*$ to $\cD^*$.

\begin{remark} \label{0245} \mar{0245}
There is a natural quasi-isomorphism $\cU^{(0)}_1$ of complex
$\cV^*$ to the complex $\cD^*$. This morphism associates to a
multivector field $\vt_0\w\cdots\w\vt_n$ the multidifferential
operator whose action on functions $f_0,\ldots,f_n\in C^\infty(Z)$
is given by the expression
\be
\frac{1}{(n+1)!}\op\sum_s \mathrm{sgn}(s)\vt_{s(0)}(f_0)\cdots
\vt_{s(n)}(f_n)
\ee
where $s$ runs through all permutations of the numbers
$(0,\ldots,n)$ and sgn$(s)$ is the sign of a permutation $s$. For
instance, $\cU^{(0)}_1$ assigns to a Poisson bivector $w$ the
Poisson bracket $\frac12\{,\}$. However, the morphism
$\cU^{(0)}_1$ fails to preserve the Lie structure.
\end{remark}

An $L_\infty$-quasi-isomorphism $\cU$ of $\cV^*$ to $\cD^*$
intertwines their differential graded Lie algebra structures and
solutions of the Maurer--Cartan equations (\ref{0205}) and
(\ref{0214}). Let us note that this $L_\infty$-quasi-isomorphism
fails to be canonical. It is represented by a power series whose
first term of differential operators of minimal order coincides
with the morphism $\cU^{(0)}_1$ in Remark \ref{0245}. This
morphism associates to each Poisson bivector field $w\in\cV^1$ on
$Z$ a certain bidifferential operator $\al_w\in \cD^1$ which obeys
the Maurer--Cartan equation (\ref{0214}) and, thus, defines a
differential associative deformation ${\mathfrak m} +\al_w$ of the
ring $C^\infty(Z)$.

An $L_\infty$-quasi-isomorphism $\cU_1$ of $\cV^1$ to $\cD^1$ on
$Z=\mathbb R^r$ in the explicit form has been obtained in
\cite{konts0,konts2}.

\section{Deformation quantization on jet manifolds}

Let $X$ be a smooth manifold of dimension $n>1$ and $C^\infty(X)$
a real ring of real smooth functions on $X$. As was mentioned
above, we restrict our study of deformation quantization on $X$ to
differential deformations whose terms of non-zero degree in a
deformation parameters are multidifferential operators.

As was mentioned above, jet formalism provides the adequate
formulation of theory of differential operators and differential
equations on manifolds and fibre bundles \cite{bryant,book,kras}.
Therefore, we develop an idea in \cite{cat02,cat04,book05} and aim
to describe deformation quantization in terms of jets.

Since deformation terms are multilinear differential operators,
but not the linear ones, you follow the technique of jets of
sections of fibre bundles over $X$ \cite{book,book13,sau}. Note
that jets of sections are the particular jets of maps
\cite{kol,rei}. If these are sections of vector bundles, their
jets coincide with jets of $C^\infty(X)$-modules
\cite{kras,book00} which are representative objects of linear
differential operators on modules \cite{kras,book12}.

Following Definition \ref{df42} of differential operators, we
formulate multidifferential operators in the jet terms (Definition
\ref{df73}) and define their infinite order jet prolongation on
the infinite order jet manifold $J^\infty F$ (\ref{df105}) of jets
of smooth real functions on a manifold $X$.

\subsection{Multidifferential operators on $C^\infty(X)$}

In the framework of formalism of jets of sections of fibre bundles
(Appendix), a real vector space $C^\infty(X)$ of smooth real
functions on a manifold $X$ is represented as a structure module
$F(X)$ of global sections of a trivial vector bundle
\mar{df20}\beq
F=X\times\mathbb R\to X, \label{df20}
\eeq
provided with bundle coordinates $(x^\m,t)$ where a coordinate $t$
possesses the identity transition functions, i.e., it is a global
coordinate on $\mathbb R$. Without the loss of generality, we
assume that it is the Cartesian coordinate on the linear space of
real number $\mathbb R$.

The tangent bundle, vertical tangent bundle and the cotangent
bundle of the fibre bundle $F$ (\ref{df20}) are respectively  the
following:
\mar{df21,2,3}\ben
&& TF=TX\times\mathbb R\times \mathbb R= TX\op\times_X
F\op\times_X F, \label{df21}\\
&& VF=F\times \mathbb R=F\op\times F, \label{df22}\\
&& T^*F=T^*X \times\mathbb R\times \mathbb R= T^*X\op\times_X
F\op\times_X F. \label{df23}
\een

Let $J^1F$ be the first order jet manifold of a sections of the
fibre bundle $F$ (\ref{df20}). It is provided with the adapted
coordinates $(x^\la, t, t_\la)$ (\ref{50}) possessing transition
functions
\mar{df24}\beq
t'_\la = \frac{\dr x^\m}{\dr{x'}^\la}t_\m. \label{df24}
\eeq
Comparing the  expressions (\ref{df24}) and (\ref{df27}) shows
that the first order jet manifold $J^1F$ is isomorphic to a bundle
product
\mar{df25}\beq
J^1F=F\op\times_X T^*X. \label{df25}
\eeq
Herewith, the jet bundles (\ref{1.14}):
\mar{df29}\beq
\pi^1:J^1F\to X, \qquad \pi^1_0:J^1F\to F, \label{df29}
\eeq
are vector bundles.

Accordingly, the canonical imbeddings (\ref{18}) -- (\ref{24})
take a form
\mar{df30,1}\ben
&&\la_1:J^1F\op\to_F
T^*X \op\otimes_F TF, \quad \la_1=dx^\la \otimes (\dr_\la + t_\la
\dr_t)=dx^\la\otimes
d_\la, \label{df30}\\
&&\thh_1:J^1F \to T^*F\op\otimes_F VF, \quad
 \thh_1=(dt- t_\la dx^\la)\otimes \dr_t=\thh \otimes
\dr_t,\label{df31}
\een
where $d_\la$ are total derivatives and $\thh$ is the contact
forms.  Identifying the jet manifold $J^1F$ to its images under
the canonical morphisms (\ref{df30}) and (\ref{df31}), one can
represent jets $j^1_xs=(x^\la,t,t_\m)$ by tangent-valued forms
\mar{df32}\beq
dx^\la \otimes (\dr_t + t_\la \dr_y), \qquad (dt- t_\la
dx^\la)\otimes\dr_t. \label{df32}
\eeq

Let $J^rF$ now be the $r$-order jet manifold of the fibre bundle
$F$ (\ref{df20}). It is provided with the adapted coordinates
$(x^\la,t_\La)$ (\ref{55.1}) possessing transition functions
(\ref{55.21}):
\mar{df33}\beq
t'_{\la+\La}=\frac{\dr x^\m}{\dr x'^\la}d_\m t'_\La. \label{df33}
\eeq
where the total derivative $d_\la$ read
\mar{df34}\beq
d_\la = \dr_\la + \op\sum_{0\leq|\La|} t_{\La+\la}\frac{\dr}{\dr
t_\La}. \label{df34}
\eeq

Finite order jet manifolds $J^rF$ constitute the inverse system
(\ref{5.10}):
\mar{df35}\beq
X\op\longleftarrow^\pi F\op\longleftarrow^{\pi^1_0}\cdots
\longleftarrow J^{r-1}F \op\longleftarrow^{\pi^r_{r-1}}
J^rF\longleftarrow\cdots, \label{df35}
\eeq
whose projective limit (\ref{df106}) is the infinite order jet
manifold
\mar{df105}\beq
J^\infty F=\op\lim^\lto J^rF, \label{df105}
\eeq
together with surjections $\pi^\infty$, $\pi^\infty_0$ and
$\pi^\infty_k$ (\ref{5.74}). It is a paracompact Fr\'echet
manifold. A manifold atlas $\{U,x^\la\}$ of $X$ yields the
manifold coordinate atlas
\be
\{(\pi^\infty)^{-1}(U), (x^\la, t_\La)\}, \qquad 0\leq|\La|,
\ee
of $J^\infty F$, together with the transition functions
(\ref{df33}) where $d_\la$ is the total derivative (\ref{df34}).

\begin{remark} \label{df50} \mar{df50}
One can think of $J^\infty F$ as being a space of infinite order
jets of real smooth functions on a manifold $X$.
\end{remark}

The inverse system (\ref{df35}) yields the direct system
(\ref{5.7}) of algebras of exterior form on finite order jet
manifolds and, in particular, the direct system (\ref{df38}):
\mar{df40}\beq
C^\infty(X)\op\longrightarrow^{\pi^*} C^\infty(F)
\op\longrightarrow^{\pi^1_0{}^*} C^\infty(J^1F) \longrightarrow
\cdots \longrightarrow C^\infty(J^rF) \longrightarrow\cdots
\label{df40}
\eeq
of vector spaces $C^\infty(J^rF)=\cO_r^0$ of smooth real functions
on finite order jet manifolds. Let
\mar{df45}\beq
C^\infty(J^\infty F)=\cO^0_\infty \label{df45}
\eeq
be its direct limit. It consists of smooth real functions on
finite order jet manifolds modulo the pull-back identification.

\begin{definition} \label{df250} \mar{df250}
 We agree to call elements of the direct limit (\ref{df45})
the smooth real functions of finite jet order on the infinite
order jet space $J^\infty F$.
\end{definition}

Turn now to the notion of a $k$-order $C^\infty(X)$-valued
differential operator on a real vector space $C^\infty(X)$. In
accordance with Definition \ref{ch538} it defines as a section of
the pull-back bundle (\ref{5.113}):
\mar{df41}\beq
J^kF\op\times_X F =J^kF\times \mathbb R\to J^kF, \label{df41}
\eeq
(cf. (\ref{df20})) which is smooth real function on the jet
manifold $J^kF$. Thus, we come to the following.

\begin{definition} \label{df42} \mar{df42}
The $k$-order $C^\infty(X)$-valued differential operators on a
real vector space $C^\infty(X)$ are given by smooth real functions
$\cE$ on the jet manifold $J^kF$.
\end{definition}

Furthermore, by very definition of the direct limit (\ref{df45})
of the direct system (\ref{df40}), there is a pull-back
monomorphism
\mar{df43}\beq
\pi^\infty_k{}^*: C^\infty(J^kF)\to C^\infty(J^\infty F).
\label{df43}
\eeq
Consequently, we come to the following.

\begin{theorem} \label{df251} \mar{df251}
Any finite order $C^\infty(X)$-valued differential operator on
$C^\infty(X)$ (Definition \ref{df42}) is represented by the
pull-back smooth real function
\mar{df70}\beq
\pi^\infty_k{}^*\cE\in C^\infty(J^\infty F) \label{df70}
\eeq
of finite jet order on the infinite order jet manifold $J^\infty
Y$.
\end{theorem}

Let us generalize this construction to multidifferential
operators.

\begin{definition} \label{df73} \mar{df73}
A $C^\infty(X)$-valued $m$-multidifferential operator on
$C^\infty(X)$ is defined to be a $m$-linear smooth real function
of finite jet order on some product of the infinite order jet
space
\mar{df71}\beq
\Pi^m=\op\times_X^m J^\infty F, \label{df71}
\eeq
which is the pull-back of a $m$-linear smooth real function on a
product of finite order jet spaces
\mar{df72}\beq
\op\times_X^m J^* F=J^{k_1}F\op\times_X\cdots \op\times_X J^{k_m}
F. \label{df72}
\eeq
\end{definition}

\begin{example} \label{df96} \mar{df96}
Any multivector field $\vt$ (\ref{cc6}) on a manifold $X$ yields a
multidifferential operator on $C^\infty(X)$ defined by a function
\mar{df99}\beq
\vt=\frac1m \vt^{\la_1\ldots\la_m}t^1_{\la_1}\cdots t^m_{\la_m}
\label{df99}
\eeq
on the product $\Pi^m$ (\ref{df71}). In these terms, the
Schouten--Nijenhuis bracket (\ref{z30}) reads
\mar{df271}\ben
&& [\vt,\up]_\mathrm{SN} =\vt\bullet\up
+(-1)^{ms}\up\bullet\vt, \label{df271} \\
&& \vt\bullet\up = \frac{m}{m!s!}(\vt^{\mu\la_2\ldots\la_m}
d_\m\up^{\al_1\ldots\al_s},t^1_{\la_2},
\ldots,t^{m-1}_{\la_m},t^m_{\al_1},\ldots,t^{m+s-1}_{\al_s}).
\nonumber
\een
\end{example}

\begin{example} \label{df100} \mar{df100}
Let $(X,w)$ be a Poisson manifold. Then the Poisson bracket
$\{,\}$ (\ref{b450}) is a bidifferential operator on $C^\infty(X)$
given by a function
\mar{df101}\beq
w=\frac12 w^{\mu\nu}t^1_\mu t^2_\nu, \qquad w^{\mu\la_1}\dr_\mu
w^{\la_2\la_3}\dr_{\la_1}\w\dr_{\la_2}\w\dr_{\la_3}=0,
\label{df101}
\eeq
on $\Pi^2$.
\end{example}

Considering $C^\infty(X)$-valued multidifferential operators, one
can say something more. There are both a canonical bundle
isomorphism
\mar{df146}\beq
J^k(\op\times_X^m F)=\op\times_X^m J^kF \label{df146}
\eeq
and a fibration
\mar{df145}\beq
\op\times_X^mJ^{\mathrm{max}(k_i)}F\to J^{k_1}F\op\times_X\cdots
\op\times_X J^{k_m} F. \label{df145}
\eeq
Any smooth real function on the product (\ref{df72}) yields the
pull-back function on the product (\ref{df145}) and, consequently,
on the jet manifold (\ref{df146}). This function gives rise again
to the pull-back function on
\mar{df147}\beq
\Pi^m=J^\infty(\op\times_X^m F)\cong\op\times_X^m J^\infty F.
\label{df147}
\eeq
Thus, we come to the following.

\begin{theorem} \label{df148} \mar{df148}
Any $C^\infty(X)$-valued $m$-multilinear $k$-order differential
operator on $C^\infty(X)$ is defined by a $m$-linear smooth real
function on the $k$-order jet manifold (\ref{df145}), and it is
represented by the pull-back function
\be
\f^m(x,t^1,\ldots, t^m)=\f^{\La_1\ldots\La_m}(x)t^1_{\La_1},\cdots
t^m_{\La_m}, \qquad |\La_i|\leq k,
\ee
of finite jet order on the infinite
order jet manifold (\ref{df147}).
\end{theorem}

Let $D^m\subset C^\infty(\Pi^m)$ denote a set of these functions.
In accordance with Theorem \ref{df148}, there is one-to-one
correspondence between $D^m$ and the set $\cD^m$ (\ref{df288}) of
multidifferential operators on $C^\infty(X)$.  By analogy with the
subcomplex $\cD^*$ of the Hochschild complex
$B^*(C^\infty(X),C^\infty(X))$ (\ref{ws50}), let us define the
Hochschild complex
\mar{df255,6}\ben
&& D^1 \ar^{\dl^1} D^2\ar^{\dl^2}\cdots\ar
D^m\ar^{\dl^m}\cdots,
\label{df255} \\
&&
(\dl^m\f^m)(x,t^1,\ldots, t^{m+1})= t_1\f^m(t_2,\ldots,t^{m+1}) +
\label{df256}\\
&& \quad  \op\sum_j(-1)^j\f^m
(t^1,\ldots,t^jat^{j+1},\ldots,t^{m+1}) +
(-1)^{m+1}\f^m(t^1,\ldots,t^m)t^{m+1}. \nonumber
\een
For instance, we have
\mar{df260,1}\ben
&& \dl^1f^1(t,t')=tf^1(t')-f^1(tt') +f^1(t)t', \label{df260}\\
&& \dl^2 f^2(t,t',t'')=  tf^2(t',t'') - f^2(tt',t'')
+  \label{df261}\\
&& \qquad f^2(t,t't'') - f^2(t,t')t''. \nonumber
\een
The complex $D^*$ (\ref{df255}) is provided with the composition
product (\ref{ws70}):
\mar{df257}\ben
&& \f^m\circ \f^n(t^1,\ldots,t^{m+n-1})= \label{df257}\\
&&\quad \op\sum_{i=1}^m (-1)^{(i-1)(n-1)}
\f^m(t^1,\ldots,t^{i-1},\f^n(t^i,\ldots, t^{n+i-1}),
t^{n+i},\ldots, t^{m+n-1}), \nonumber
\een
and the Gerstenhaber bracket (\ref{ws73}):
\mar{df258}\beq
[\f,\f']_\mathrm{G}=\f\circ \f' -(-1)^{|\f||\f'|}\f'\circ \f.
\label{df258}
\eeq

One can show that the bracket (\ref{df258}) obeys the graded
Jacobi identity (\ref{0211}). Thus, $(D^*,[.,.]_\mathrm{G})$ is a
graded Lie algebra. Furthermore, let us consider a one-cocycle
\mar{df280}\beq
\f^1(t,t')={\mathfrak m}(t,t')=tt', \label{df280}
\eeq
and introduce the operator
\mar{df265}\beq
d_{\mathfrak m} \f=[{\mathfrak m},\f]_\mathrm{G} \label{df265}
\eeq
of degree one on $D^*$. A direct computation shows that
\be
d_{\mathfrak m}\f=(-1)^{(|\f|-1)}\dl^{|\f|}\f,
\ee
and the relation (\ref{0212}) holds. The bracket (\ref{df258}) and
the differential (\ref{df265}) bring the complex $D^*$
(\ref{df255}) into a DGLA, isomorphic to the DGLA $\cC^*$
(\ref{0206}). Let $H^*(D^*)$ denote its cohomology.

As was mentioned above, any multivector field $\vt\in \cT_*(X)$
(\ref{cc6}) on a manifold $X$ yields the multidifferential
operator $\vt$ (\ref{df99}) (Example \ref{df96}). Let $\cT^*$
denote its subset in $D^*$ (Remark \ref{0245}). It is readily
observed that
\mar{df270}\beq
d_{\mathfrak m}\vt=0 \label{df270}
\eeq
for any multivector field $\vt$ (\ref{df99}). It follows that
$\cT^*$ is a subcomplex of the Hochschild complex $D^*$
(\ref{df255}) whose coboundary operators equal zero. It follows
that there is a natural bijection between cohomology $H^*(\cT^*)$
of this complex and $\cT^*$ itself. The modification
$[.,.]_\mathrm{SN'}$ (\ref{gm5}) of the Schouten--Nijenhuis
bracket (\ref{df271}) and the coboundary operator $d_{\mathfrak
m}$ (\ref{df270}) bring the complex $\cT^*$ into a DGLA,
isomorphic to the DGLA $\cV^*$ (\ref{0220}). The cohomology
$H^*(\cT^*)$ of a DGLA $\cT^*$ also is a DGLA (Theorem
\ref{df272}). As we have shown above, there is an isomorphism of
DGLAs $\cT^*=H^*(\cT^*)$. Consequently, a DLA $\cT^*$ is formal
(Definition \ref{df276}).

Herewith, it should be emphasized that a monomorphism of cochain
complexes $\cT^*\to D^*$ (cf. (\ref{0124})) fails to be a
morphisms of DGLAs because the Schouten--Nijenhuis bracket
(\ref{df271}) in $\cT^*$ is not induced by the Gerstenhaber
bracket (\ref{df258}) in $D^*$ (cf. Remark \ref{0245}). At the
same time, Theorem \ref{0240} states that there exists a
quasi-isomorphism (Definition \ref{df275})
\mar{df277}\beq
D^*\to\cT^*=H^*(\cT^*), \label{df277}
\eeq
that is, we have an isomorphism
\mar{df278}\beq
H^*(D^*)=H^*(\cT^*)=\cT^* \label{df278}
\eeq
(cf. (\ref{0126})). In accordance with Theorem \ref{0222}, a DGLA
$D^*$ is formal, i.e., there exist both the quasi-isomorphism
(\ref{df277}) of a DGLA $\cD^*$ to $H^*(D^*)= \cT^*$ and the
inverse $L_\infty$-quasi-isomorphism $\cT^*\to D^*$.

\subsection{Deformations of $C^\infty(X)$}

Turn now to deformations of a real ring $C^\infty(X)$. In order to
describe them in terms of jets of functions, let us treat a
multiplication
\be
(f,f')\to ff', \qquad f,f'\in C^\infty(X),
\ee
in $C^\infty(X)$ as a bidifferential operator of zero order. In
accordance with Definition \ref{df73}, it is represented by a real
bilinear function (\ref{df280}):
\mar{df75}\beq
\cE_0(x^\m,t,t_\La,t',t'_\La)=\mathrm{m}(t,t')=tt', \label{df75}
\eeq
on the product $\Pi^2$ (\ref{df71}). It is the pull-back of a real
smooth bilinear function
\mar{df80}\beq
\cE_0(x^\m,t,t')=tt' \label{df80}
\eeq
on the product $F\times_XF$ (\ref{df72}).

Given a deformation parameter $h$, let $\mathbb R[[h]]$ be a real
ring of power series in $h$. It is defined as follows. Let
$\mathbb R[h]$ be a real commutative ring of polynomials in a
quantity $h$. Let $h^k\mathbb R[h]$, $k=1,2,\ldots,$ denote its
two-sided ideal generated by an element $h^k\in \mathbb R[h]$.
Then the quotient
\mar{df160}\beq
\mathbb R[h]^{k-1}=\mathbb R[h]/h^k\mathbb R[h] \label{df160}
\eeq
is a real commutative ring, whose ring space is a smooth manifold
$\mathbb R^k$. In particular, $\mathbb R[h]^0=\mathbb R$ (Remark
\ref{df2}).

The ring (\ref{df160}) also is the quotient
\be
\mathbb R[h]^{k-1}=\mathbb R[h]^{k+r}/h^k\mathbb R[h]^{k+r},
\ee
and we have a ring epimorphism
\mar{df163}\beq
\xi^{k+1}_k: \mathbb R[h]^{k+1}\to \mathbb R[h]^k, \qquad
k\in\mathbb N. \label{df163}
\eeq
This epimorphism also is a fibration of ring manifolds $\mathbb
R^{k+2}\to \mathbb R^{k+1}$.

The rings $\mathbb R[h]^k$ (\ref{df160}) together with
epimorphisms $\xi^{k+1}_k$ constitute the inverse system of real
rings
\mar{df165}\beq
\mathbb R \longleftarrow \mathbb R[h]^1 \longleftarrow\cdots
\longleftarrow \mathbb R[h]^k \longleftarrow \cdots. \label{df165}
\eeq
Its inverse image is the above mentioned real ring $\mathbb
R[[h]]$ of power series in $h$ together with epimorphisms
\mar{df166}\beq
\xi^\infty_k: \mathbb R[[h]]\to \mathbb R[h]^k, \qquad
\xi^\infty_r=\xi^k_r\circ \xi^\infty_k. \label{df166}
\eeq
In particular,
\mar{df184}\beq
\xi^\infty_0: \mathbb R[[h]]\to \mathbb R[h]^0. \label{df184}
\eeq

Since the inverse system (\ref{df165}) also is an inverse system
of smooth fibre bundles over a point, its inverse image $\mathbb
R[[h]]$ is provided with the inverse limit topology. This is the
coarsest topology such that the surjections $\xi^\infty_k$
(\ref{df166}) are continuous. The base of open sets of this
topology in $\mathbb R[[h]]$ consists of the inverse images of
$\mathbb R^{k+1}$, $k\in\mathbb N$, under the mappings
(\ref{df166}). This topology makes $\mathbb R[[h]]$ into a
paracompact Fr\'echet manifold $\mathbb R^\infty$ possessing
global manifold coordinates $(t^k, k\in\mathbb N)$.

A $C^\infty(X)$-ring $C^\infty(X)[[h]]$ of power series in $h$ is
defined as a tensor product over $\mathbb R$ of real rings
$C^\infty(X)$ and $\mathbb R[[h]]$:
\mar{df161}\beq
C^\infty(X)[[h]]=C^\infty(X)\op\ot_{\mathbb R}\mathbb R[[h]].
\label{df161}
\eeq
It is endowed with product
\be
(f\ot s)(f'\ot s')=(ff')\ot (ss'),
\ee
together with the canonical isomorphisms
\mar{df162,'}\ben
&& \mathbb R\op\ot_{\mathbb R}\mathbb R[[h]]=\mathbb R[[h]], \label{df162}\\
&& C^\infty(X)\op\ot_{\mathbb R} \mathbb R=C^\infty(X). \label{df162'}
\een
The isomorphism (\ref{df162}) provides $C^\infty(X)[[h]]$ with the
structure of $\mathbb R[[h]]$-ring, whereas the isomorphism
(\ref{df162'}) yields the canonical monomorphism
\mar{df170}\beq
C^\infty(X)\to C^\infty(X)\op\ot_{\mathbb R} \bb=
C^\infty(X)\op\ot_{\mathbb R} \mathbb R \subset C^\infty(X)[[h]].
\label{df170}
\eeq
We have also an epimorphism of real rings
\mar{df183}\beq
\xi_0=\id\ot\xi^\infty_0: C^\infty(X)[[h]]\to
C^\infty(X)\op\ot_{\mathbb R} = C^\infty(X), \label{df183}
\eeq
where $\xi^\infty_0$ is the epimorphism (\ref{df184}).

In order to define multidifferential operators on
$C^\infty(X)[[h]]$ let us start with $\mathbb R[[h]]$-linear
morphisms
\mar{df172}\beq
\phi: C^\infty(X)[[h]]\to C^\infty(X)[[h]] \label{df172}
\eeq
of an $\mathbb R[[h]]$-module $C^\infty(X)[[h]]$. Obviously, such
a morphism $\phi$ is defined in full by its restriction to the
subring $C^\infty(X)$ (\ref{df170}) of $C^\infty(X)[[h]]$. This
restriction reads
\mar{df171}\beq
\phi(f)=\op\sum_{r\in \mathbb N}h^r\phi_r(f), \qquad f\in
C^\infty(X), \label{df171}
\eeq
where $\phi_k$ are $\mathbb R$-linear morphisms of a real vector
space $C^\infty(X)$. The morphism $\phi$ (\ref{df172}) is said to
be an automorphism of $C^\infty(X)[[h]]$ if it is its $\mathbb
R[[h]]$-isomorphism.

If $\phi$ (\ref{df172}) is an automorphism of $C^\infty(X)[[h]]$,
then $\phi_0=\xi_0\circ \phi$ is an automorphism of a real vector
space $C^\infty(X)$. Such an automorphism is a multiplication
\mar{df190}\beq
\phi_0(f)=gf, \qquad f\in C^\infty(X), \label{df190}
\eeq
where $g$ is an invertible element of $C^\infty(X)$. Then any
automorphism $\phi$ (\ref{df172}) of $C^\infty(X)[[h]]$ is a
composition of a multiplication
\mar{df191}\beq
\phi_g:f_h\to gf_h, \qquad f_h\in C^\infty(X)[[h]], \label{df191}
\eeq
by an invertible element $g\in C^\infty(X)$ and an automorphism
$\phi$ so that $\phi_0$ (\ref{df190}) is the identity map. The
latter takes the following form (\ref{df171}):
\mar{df192}\beq
\phi(f)=(1 +h\si)f, \qquad f\in C^\infty(X), \label{df192}
\eeq
where $\si$ is an arbitrary $\mathbb R[[h]]$-linear, but not
$C^\infty(X)$-linear endomorphism of $C^\infty(X)[[h]]$. The
inverse of $\phi$ (\ref{df192}) is
\be
\phi^{-1}(f)= \left(1 + \op\sum_k (-1)^kh^k\op\circ^k\si\right)f,
\ee
where $\circ$ denotes the composition of morphisms. Thus, a
generic automorphism of $C^\infty(X)[[h]]$ is given by the
expression
\mar{df194}\beq
\phi(f_h)=g(1+h\si)f_h, \qquad f_h\in C^\infty(X)[[h]],
\label{df194}
\eeq
where $g$ is an invertible element of $C^\infty(X)$. Note that it
also is a $C^\infty(X)$-linear morphism of $C^\infty(X)[[h]]$.

 We say that the $\mathbb
R[[h]]$-linear morphism $\phi$ (\ref{df172}) is a differential
operator if it is a finite order linear differential operator on a
$\mathbb R[[h]]$-module $C^\infty(X)[[h]]$. In this case, each
morphism $\phi_k$ in the expression (\ref{df171}) is a finite
order $C^\infty(X)$-valued differential operator on $C^\infty(X)$.

In accordance with Definition \ref{df42}, a differential operator
$\phi_r$ is represented by a smooth real function on some finite
order jet manifold $J^kY$ or its pull-back onto the infinite order
jet manifold $J^\infty F$. Then one can regard the morphism as
$\mathbb R[[h]]$-valued function on $J^\infty F$ whose components
$\phi_r$ are the above-mentioned pull-back of functions on finite
order jet manifolds.

In view of Theorem \ref{df148}, we then can define a
$C^\infty(X)[[h]]$-valued $m$-linear differential operator $\ve$
on $C^\infty(X)[[h]]$ as follows.

\begin{definition} \label{df176} \mar{df176}  A
$C^\infty(X)[[h]]$-valued $m$-multidifferential operator $\ve$ on
an $\mathbb R[[h]]$-ring $C^\infty(X)[[h]]$ is an $\mathbb
R[[h]]$-multilinear map whose restriction on a real ring
$C^\infty(X)$ is represented by an $\mathbb R[[h]]$-valued
$m$-linear function
\mar{df177}\beq
\ve=\op\sum_{r\in \mathbb N}h^r\ve_r \label{df177}
\eeq
on the product $\Pi^m$ (\ref{df147}) such that its components
$\ve_r$ are smooth real functions on some finite order jet
manifold (\ref{df146}).
\end{definition}

It follows from this definition that, in particular, any
multidifferential operator $\cE$ on $C^\infty(X)$ (Theorem
\ref{df148}) defines the multidifferential operator (\ref{df177})
on $C^\infty(X)[[h]]$ where $\cE_{r>0}=0$.

For instance, the multiplication (\ref{df80}) in $C^\infty(X)$
also is a multiplication in $C^\infty(X)[[h]]$. At the same time,
any bidifferential operator
\mar{df180}\beq
\ve(t,t')=\op\sum_{r\in\mathbb N}h^r\ve_r(x,t_\La,t'_\Sigma)=
\op\sum_{r\in\mathbb N}h^r\ve_r^{\La,\Si}(x)t_\La t'_\Si
\label{df180}
\eeq
defines a different multiplication
\mar{df181}\beq
t*t'=\op\sum_{r\in\mathbb N}h^r\ve_r(x,t_\La,t'_\Sigma)
\label{df181}
\eeq
in $C^\infty(X)[[h]]$.

Following the standard terminology, we say that the multiplication
(\ref{df181}) is a deformation of a ring $C^\infty(X)$. In
accordance with Definition \ref{df231}, two deformations $\ve$ and
$\ve'$ of $C^\infty(X)$ are equivalent if there exists the
automorphism $\phi$ (\ref{df172}) of $C^\infty(X)[[h]]$ so that
the relation (\ref{070}):
\mar{df200}\beq
\phi(f_h*'f'_h)=\phi(f_h)*\phi(f'_h), \qquad f_h,f'_h\in
C^\infty(X)[[h]], \label{df200}
\eeq
holds.

Now let us require that the deformation (\ref{df181}) is a ring,
i.e., $C^\infty(X)[[h]]$ is a unital associative algebra with
respect to this product. We prove item (i) of Lemma \ref{df9}.

Let $\bb_h$ be the unit element with respect to the product
(\ref{df181}). Referring to the expression (\ref{df181}), we
obtain
\be
f=\bb_h*f= \op\sum_{r\in\mathbb N}h^r\ve_r(x,\bb_h,f)=
\op\sum_{r\in\mathbb N}h^r\ve_r(x)\bb_hf,
\ee
for any $f\in C^\infty(X)$. It follows that
\be
\op\sum_{r\in\mathbb N}h^r\ve_r(x)\bb_h=1,
\ee
that is, $\bb_h$ is some invertible element of $C^\infty(X)[[h]]$.
Then there is the automorphism $f\to (\bb_h)^{-1}f$ (\ref{df194})
which bring $\bb_h$ into 1.

Hereafter, we therefore assume that the deformation (\ref{df181})
preserves 1 as the unit. In this case, the coefficient function
$\ve_r^{\La,\Si}$, $|\La|\leq 0$, $|\La|\leq 0$, in the expression
(\ref{df180}) obey the relations
\mar{df201}\beq
\ve_0^{.,.}(x)=1, \qquad \ve_{r>0}^{.,\Si}(x)=\ve_{r>0}^{\La,.}=0,
\label{df201}
\eeq
i.e., the bidifferential operators $\ve_{r>0}$ do not contain
zero-order differential operators. Such a deformation takes a form
\mar{df202}\ben
&& t*t'=tt'+\op\sum_{0<r, 0<|\La|,
0<|\Si|}h^r\ve_r(x,t_\La,t'_\Sigma)=
\label{df202}\\
&& \qquad tt'+\op\sum_{0<r, 0<|\La|,
0<|\Si|}h^r\ve_r^{\La,\Si}(x)t_\La t'_\Si. \nonumber
\een

We also restrict our consideration to automorphisms (\ref{df192})
of $C^\infty(X)[[h]]$. Since we deal with differential
deformations, such an automorphism is represented by a linear
bundle morphism
\mar{df210}\ben
&& \f: J^\infty F\op\to_X F, \label{df210}\\
&&  t'=t+h\f(x,t_\La)=t +h\op\sum_{k\in\mathbb N} h^k\si_k(x, t_\La), \qquad
0\leq|\La|,\nonumber
\een
where $\si_k$ are functions on $J^\infty F$ (i.e., linear
differential operators on $C^\infty(X)$) of finite jet order which
are independent of $t$ and linear in $t_\La$, $0<|\La|$. In this
case, the condition (\ref{df200}) of an equivalence of different
deformations $\ve$ and $\ve'$ (\ref{df202}) takes a form
\mar{df211}\ben
&& tt'+ h(t\si(x,t'_\La)+\si(x,t_\La)t') + \label{df211} \\
&& \qquad \op\sum_{0<r, 0<|\La|, 0<|\Si|}h^r\ve_r(x,d_\La(t+h\si(x,t_\Xi)),
d_\Sigma(t'+h\si(x,t'_\Xi)))= \nonumber \\
&& \qquad \phi(tt') +\phi\left(\op\sum_{0<r, 0<|\La|,
0<|\Si|}h^r\ve'_r(x,t_\La,t'_\Sigma)\right). \nonumber
\een
In particular,
\mar{df213}\beq
t\si_0(x,t'_\la) + \si_0(x,t_\La)t' +\ve_1(x,t,t')= \si_0(x,tt')
+\ve'_1(x,t,t'). \label{df213}
\eeq
For instance, a deformation $\ve'$ is trivial only if
\mar{df212}\beq
\ve'_1(x,t,t,)=t\si_0(x,t'_\La) + \si_0(x,t_\La)t' - \si_0(x,tt').
\label{df212}
\eeq

Let us now investigate the associativity condition (\ref{qm885})
of the deformation (\ref{df202}):
\mar{df203}\beq
(t*t')*t''-t*(t'*t'')=0.
\eeq
We come to the following relation of three-linear differential
operators
\mar{df203}\beq
D_r(x,t,t',t'')=\op\sum_{0\leq k\leq r}
(\ve_{r-k}(x,\ve_k(t,t'),t'')-\ve_{r-k}(x,t, \ve_k(t',t''))=0.
\label{df203}
\eeq
In particular,
\be
D_0(x,t,t',t'') = (tt')t''-t(t't'')=0,
\ee
and
\mar{df204}\ben
&& D_1(x,t,t',t'')= d_\mathrm{m}\ve_1(x,t,t',t'')= \label{df204} \\
&& \qquad \ve_1(x,tt',t'')-\ve_1(x,t, t't'')+
\ve_1(x,t,t')t''-t\ve_1(x,t',t'')=0. \nonumber
\een
We obtain from the relation (\ref{df204}) that
\be
\ve_1^{\La+\La',\Si}- \ve_1^{\La,\La'+\Si}=0,
\ee
where functions $\ve_1$ also obey the conditions (\ref{df201}).

Following the procedure in Section 2.1.2, let us write
\mar{df205-7}\ben
&& D_r(x,t,t',t'')=E_r(x,t,t',t'') -(\dl^2 \ve_r)(x,t,t',t''),
\label{df205}\\
&& E_r(x,t,t',t'')= \label{df206}\\
&& \qquad \op\sum_{0<k<r} (\ve_{r-k}(x,\ve_k(t,t'),t'')-\ve_{r-k}(x,t,
\ve_k(t',t'')), \nonumber\\
&& (\dl^2 \ve_r)(x,t,t',t'')=t\ve_r(x,t',t'')-\ve_r(x,tt',t'')+
\label{df207}\\
&& \qquad \ve_r(x,t,t't'')-\ve_r(x,t,t')t'', \nonumber
\een
where $\dl^2$ (\ref{df207} is the second Hochschild coboundary
operator (\ref{df261}).

For instance, a glance at the expression (\ref{df204}) shows that
\mar{df208}\ben
&& E_1(x,t,t',t'')=0, \nonumber\\
&& D_1(x,t,t',t'')= -(\dl^2 \ve_1)(x,t,t',t'')=0, \label{df208}
\een
that is, a deformation $\ve$ is associative only if its term
$\ve_1$ is a Hochschild two-cocycle. In particular, the condition
(\ref{df208}) is satisfied if $\ve_1$ is a Hochschild coboundary,
i.e.,
\mar{df233}\beq
\ve_1(x,t,t')=(\dl^1\g)(x,t,t')=t\g(x,t')-\g(x,tt')+\g(x,t)t'
\label{df233}
\eeq
where $\dl^1$ is the first Hochschild coboundary operator
(\ref{df260}). A glance at the relation (\ref{df212}) shows that
this is the case of a trivial deformation when $\phi$ is some
automorphism (\ref{df210}) and $\g=\si_0$ (Theorem \ref{093}).

For instance, let $\g$ be a first order differential operator.
Since it must not contain zero order differential operator, it is
a derivation, and then $\ve_1$ (\ref{df233}) equals zero.

Let $\si_0$ be a second order differential operator, i.e.,
\mar{df234}\beq
\g(x,t)=\g^{\mu\nu}(x)t_{\mu\nu}, \label{df234}
\eeq
where certainly $\g^{\mu\nu}=\g^{\nu\mu}$. Then
\mar{df235}\beq
\ve_1(x,t,t')=t\g^{\mu\nu}(x)d_{\mu\nu}t' +
t'\g^{\mu\nu}(x)d_{\mu\nu}t -
\g^{\mu\nu}(x)d_{\mu\nu}(tt')=\g^{\mu\nu}t_\mu t_\nu \label{df235}
\eeq
is a Hochschild two-coboundary. Let us consider
\mar{df236}\beq
\ve_1(x,t,t')= \ve^{\mu\nu}(x)t_\mu t_\nu, \qquad
\ve^{\mu\nu}=-\ve^{\nu\mu}. \label{df236}
\eeq
Then it is readily observed that $\dl^2\ve_1(x,t,t')=0$, i.e.,
$\ve_1$ (\ref{df236}) is a Hochschild two-cocycle, but not a
coboundary.

Thus, we come to Theorem \ref{093} where $A=C^\infty(X)$ that
non-equivalent deformations of a ring $C^\infty(X)$ are
parameterized by elements of the Hochschild cohomology
$H^2(C^\infty(X),C^\infty(X))$,  and to Theorem \ref{0125} that
\be
H^2(C^\infty(X),C^\infty(X))=\cT_2(X).
\ee
Since we are restricted to differential deformations, their
equivalence classes are associated to elements of the differential
Hochschild cohomology $H^2(D^*)$. In accordance with Theorems
\ref{0240} and \ref{0222}, we have the isomorphism (\ref{0126}):
\mar{df900}\beq
H^2(D^*)=H^2(\cT^*)=\cT^2. \label{df900}
\eeq

Given a bivector field $\ve_1\in \cT^2$, a problem however lies in
that to write the corresponding deformation (\ref{df202}) in an
explicit form. If $X=\mathbb R^d$ is endowed with global
coordinates $(x^\la)$ and $\ve_1$ is given by the expression
(\ref{df236}) with constant coefficients $\ve^{\m\nu}$, the Weyl
product provides a desired deformation. It reads
\mar{df290}\beq
t* t'= \op\sum_{0\leq k} \frac{h^k}{k!}\ve^{\al_1\bt_1} \cdots
\ve^{\al_k\bt_k} t_{\al_1\ldots\al_k}t'_{\bt_1\ldots\bt_k}.
\label{df290}
\eeq

Let $w\in \cT^2$ be the Poisson bivector field (\ref{df101}), and
let $*$ be the corresponding deformation. Then the commutator
(\ref{0117}):
\mar{df301}\beq
[f,f']_h=\frac1{2h}(f* f'-f'* f) \label{df301}
\eeq
defines a Lie deformation of the Poisson bracket $\{,\}$ on $X$
(Remark \ref{df300}). For instance, let $w$ provide a regular
Poisson structure on $\mathbb R^d$ and $(q^i,p_i, z^a)$ the
corresponding Darboux coordinates such that
\mar{df800}\beq
w=\frac12(t^it'_i-t_it'^i). \label{df800}
\eeq
Then the corresponding Lie deformation is given by the Moyal
product (\ref{qm893}):
\mar{df302}\beq
[f,f']_h=\op\sum_{k=0}\frac{h^k}{k!} \op\sum_{r=0}^k
(-1)^r(t_{i_1\ldots i_r}{}^{i_{r+1}\ldots i_k} t'^{i_1\ldots
i_r}{}_{i_{r+1}\ldots i_k}). \label{df302}
\eeq

Now we aim to generalize the Weyl expression (\ref{df290}) to an
arbitrary bivector field (\ref{df236}) on an arbitrary manifold.
In contrast to Fedosov's quantization on a symplectic manifold in
a presence of a particular symplectic connection, we use the fact
that that an infinite order jet manifold $J^\infty Y$ of a fibre
bundle $Y\to X$ admits the canonical flat connection $d_H$
(\ref{df506}). A necessary step to this goal is a definition the
infinite order jet prolongation of differential operators.

\subsection{Jet prolongation of multidifferential operators}

In order to describe infinite jet order prolongation of
multidifferential operators, let us start with that of
differential operators.

Let $Y\to X$ and $E\to X$ be fibre bundles provided with bundle
coordinates $(x^\la,y^i)$ and $(x^\la, v^A)$, respectively. By
virtue of Definition \ref{ch539}, a $k$-order $E$-valued
differential operator on $Y$ is the fibre bundle morphism
(\ref{df110}):.
\mar{df117}\ben
&& \Delta: J^kY\op\to_X E, \label{df117}\\
&& v^A\circ \Delta= \Delta^A(x^\la, y^i, y^i_\la,\ldots,
y^i_{\la_k\cdots \la_1}). \nonumber
\een
This morphism admits the canonical $r$-order jet prolongation
(\ref{5.152}) to a fibre bundle morphism
\mar{df115}\beq
J^r\Delta: J^rJ^kY\op\to_X J^rE,  \label{df115}
\eeq
where we follow the notation (\ref{df113}) and (\ref{df114}).
 Restricted to $J^{r+k}Y\subset J^rJ^kY$ (\ref{df112}), the
 morphism (\ref{df115}) is an $(r+k)$-order $J^rE$-valued differential operator
\mar{df116}\ben
&& \cJ^r\Delta= J^r\Delta\circ\si_{rk}: J^{r+k}Y\op\to_X J^rE,
\label{df116}\\
&& v^A_\La\circ \cJ^r\Delta=d_\La\Delta^A, \quad |\La|\leq r,
\nonumber
\een
on $Y$ in accordance with Definition \ref{ch539}.

\begin{definition} \label{df125} \mar{df125}
The differential operator $\cJ^r\Delta$ (\ref{df116})  is called
the $r$-order prolongation of the differential operator $\Delta$
(\ref{df117}).
\end{definition}

Let $s$ be a section of a fibre bundle $Y\to X$. The differential
operator $\Delta$ (\ref{df117}) sends it onto the section
$\Delta\circ s$ (\ref{df139}) of a fibre bundle $E\to X$. Then the
$r$-order prolongation $\cJ^r\Delta$ (\ref{df116}) of $\Delta$
(\ref{df117}) sends $s$ onto an integral section of the jet bundle
$J^rE\to X$ which is the $r$-order jet prolongation
$J^r(\Delta\circ s)$ (\ref{df139}) of a section $\Delta\circ s$ of
$Y$.

Let $\cJ^{r+1}\Delta$ be the $(r+1)$-order prolongation
(\ref{df116}) of the differential operator $\Delta$ (\ref{df117}).
Then we have a commutative diagram
\mar{df118}\beq
\begin{array}{rcccl}
& J^{r+k}Y &\op\longleftarrow^{\pi^{r+k+1}_{r+k}} & J^{r+k}Y &\\
_{\cJ^r\Delta} & \put(0,10){\vector(0,-1){20}} & & \put(0,10){\vector(0,-1){20}}& _{\cJ^{r+1}\Delta} \\
& J^rE & \op\longleftarrow^{\pi^{r+1}_r} & J^{r+1}E&
\end{array} \label{df118}
\eeq
Let us consider the inverse system of jet manifolds $J^kY$
(\ref{5.10}):
\mar{df119}\beq
X\op\longleftarrow^\pi Y\op\longleftarrow^{\pi^1_0}\cdots
\longleftarrow J^{k-1}Y \op\longleftarrow^{\pi^k_{k-1}}
J^kY\longleftarrow\cdots, \label{df119}
\eeq
and the inverse system
\mar{df120}\beq
X\op\longleftarrow^\pi E\op\longleftarrow^{\pi^1_0}\cdots
\longleftarrow J^{r-1}T \op\longleftarrow^{\pi^r_{r-1}}
J^rE\longleftarrow\cdots \label{df120}
\eeq
of jet manifolds $J^rE$. Then the commutative diagrams
(\ref{df118}) are assembled into an order preserving morphism
\mar{df121}\beq
\begin{array}{cccrccrccc}
&\cdots &\longleftarrow & & J^{r+k}Y &\op\longleftarrow^{\pi^{r+k+1}_{r+k}} & & J^{r+k}Y & \longleftarrow &\cdots\\
& & & _{\cJ^r\Delta} & \put(0,10){\vector(0,-1){20}} & & _{\cJ^{r+1}\Delta} & \put(0,10){\vector(0,-1){20}} & & \\
&\cdots &\longleftarrow  & & J^rE &
\op\longleftarrow^{\pi^{r+1}_r} & & J^{r+1}E &\longleftarrow &
\cdots
\end{array} \label{df121}
\eeq
of the inverse systems (\ref{df119}) and (\ref{df120}). Let
$J^\infty Y$ and $J^\infty E$ be the inverse limits (\ref{df106})
of the inverse systems (\ref{df119}) and (\ref{df120}),
respectively. Then there is a morphism
\mar{df122}\beq
\cJ^\infty\Delta=\op\lim^\lto \cJ^r\Delta : J^\infty Y\to J^\infty
E \label{df122}
\eeq
so that the diagrams
\mar{df123}\beq
\begin{array}{rcccl}
& J^{r+k}Y &\op\longleftarrow^{\pi^\infty_{r+k}} & J^\infty Y &\\
_{\cJ^r\Delta} & \put(0,10){\vector(0,-1){20}} & & \put(0,10){\vector(0,-1){20}}& _{\cJ^\infty\Delta} \\
& J^rE & \op\longleftarrow^{\pi^\infty_r} & J^\infty E&
\end{array} \label{df123}
\eeq
are commutative for any $r\geq 0$.

\begin{definition} \label{df126} \mar{df126}
The morphism $\cJ^\infty\Delta$ (\ref{df122})  is called the
infinite order prolongation of the differential operator $\Delta$
(\ref{df117}).
\end{definition}

Given the coordinate atlas $(x^\la,y^i_\La)$ (\ref{jet1}) of
$J^\infty Y$ and the coordinate atlas $(x^\la,v^A_\La)$
(\ref{jet1}) of $J^\infty E$, the morphism $\cJ^\infty\Delta$
(\ref{df122}) reads
\mar{df127}\beq
v^A_\La\circ \cJ^\infty\Delta=d_\La\Delta^A, \qquad 0\leq|\La|.
\label{df127}
\eeq

Now let $F=X\times \mathbb R$ be the fibre bundle (\ref{df20})
endowed with bundle coordinates $(x^\la,t)$. In accordance with
Definition \ref{df42}, a $k$-order $C^\infty(X)$-valued
differential operator $\cE$ on a real vector space $C^\infty(X)$
is defined to be a smooth real function $\cE$ on the jet manifold
$J^kF$ , i.e., a section of the pull-back bundle (\ref{df41}).

At the same time, basing on Definition \ref{ch539} of differential
operators, we come to another equivalent definition of $\cE$.

\begin{definition} \label{df130} \mar{df130}
A $k$-order $C^\infty(X)$-valued differential operator on a real
vector space $C^\infty(X)$ is a fibre bundle morphism
\mar{df131}\ben
&& \cE: J^kF\op\to_X F, \label{df131}\\
&& t'\circ \cE=\cE(x^\la, t, t_\la,\ldots, t_{\la_k\cdots
\la_1}). \nonumber
\een
\end{definition}

In accordance with Definition \ref{df125}, an $r$-order
prolongation of the differential operator $\cE$ (\ref{df131}) is
the morphism
\mar{df132}\ben
&& \cJ^r\cE: J^{r+k}F\op\to_X J^rF, \label{df132} \\
&& t'_\La\circ \cJ^r\cE=d_\La\cE, \qquad |\La|\leq r. \nonumber
\een

The differential operator $\cJ^r\cE$ sends a section $f$ of the
fibre bundle (\ref{df20}) (i.e., a smooth real function $f$ on
$X$) onto an integral section of the jet bundle $J^rF\to X$ which
is the $r$-order jet prolongation $J^r(\cE\circ f)$ (\ref{df139})
of a function $\cE\circ f$.

Let $\cJ^{r+1}\cE$ be the $(r+1)$-order prolongation (\ref{df132})
of the differential operator $\cE$ (\ref{df131}). Then we have the
commutative diagram (\ref{df118}):
\be
\begin{array}{rcccl}
& J^{r+k}F &\op\longleftarrow^{\pi^{r+k+1}_{r+k}} & J^{r+k}F &\\
_{\cJ^r\cE} & \put(0,10){\vector(0,-1){20}} & & \put(0,10){\vector(0,-1){20}}& _{\cJ^{r+1}\cE} \\
& J^rF & \op\longleftarrow^{\pi^{r+1}_r} & J^{r+1}F&
\end{array}
\ee
These diagrams are assembled into an endomorphism
\mar{df133}\beq
\begin{array}{cccrccrccc}
&\cdots &\longleftarrow & & J^{r+k}F &\op\longleftarrow^{\pi^{r+k+1}_{r+k}} & & J^{r+k}F & \longleftarrow &\cdots\\
& & & _{\cJ^r\cE} & \put(0,10){\vector(0,-1){20}} & & _{\cJ^{r+1}\cE} & \put(0,10){\vector(0,-1){20}} & & \\
&\cdots &\longleftarrow  & & J^rF &
\op\longleftarrow^{\pi^{r+1}_r} & & J^{r+1}F &\longleftarrow &
\cdots
\end{array} \label{df133}
\eeq
of the inverse system (\ref{df35}). This endomorphism
(\ref{df133}) of the inverse system (\ref{df35}) yields a morphism
\mar{df134}\ben
&& \cJ^\infty\cE=\op\lim^\lto \cJ^r\cE : J^\infty F\to J^\infty F,
\label{df134}\\
&& t'_\La\circ \cJ^\infty\cE=d_\La\cE, \qquad 0\leq|\La|, \nonumber
\een
of its inverse limit $J^\infty F$ (\ref{df105}) so that the
diagrams
\be
\begin{array}{rcccl}
& J^{r+k}F &\op\longleftarrow^{\pi^\infty_{r+k}} & J^\infty F &\\
_{\cJ^r\cE} & \put(0,10){\vector(0,-1){20}} & & \put(0,10){\vector(0,-1){20}}& _{\cJ^\infty\cE} \\
& J^rF & \op\longleftarrow^{\pi^\infty_r} & J^\infty F&
\end{array}
\ee
are commutative for any $r\geq 0$. Following Definition
\ref{df126}, one can think of the morphism (\ref{df134}) as being
the infinite order prolongation of the differential operator
(\ref{df131}).

Let $f$ be a section of the fibre bundle $F\to X$ (\ref{df20})
(i.e., a smooth real function on $X$). Its $r$-order jet
prolongation $J^rf$ (\ref{df138}) is an integrable section of the
jet bundle $J^rF\to X$. There is a map
\mar{df140}\ben
&& J^\infty f:X\to J^\infty F, \label{df140}\\
&& t_\La\circ J^\infty f =\dr_\La f, \qquad |\La|\geq 0, \nonumber
\een
so that
\mar{df142}\beq
J^rf=\pi^\infty_r\circ J^\infty f \label{df142}
\eeq
for any $r\geq 0$. One calls $J^\infty f$ (\ref{df140}) the
infinite order jet prolongation of a function $f$.

In particular, given a function $f$ and its image $\cE\circ f$
with respect to a differential operator $\cE$, let
$J^\infty(\cE\circ f)$ be its infinite order jet prolongation.
Then the morphism $\cJ^\infty\cE$ (\ref{df134}) sends $f$ onto
$J^\infty(\cE\circ f)$ so that the relations (\ref{df142}):
\be
\cJ^r\cE\circ f=\pi^\infty_{r+k}\circ \cJ^\infty\cE\circ f,
\ee
are satisfied.

Turn now to jet prolongations of multidifferential operators. By
virtue of Theorem \ref{df148}, a $C^\infty(X)$-valued
$m$-multilinear $k$-order differential operator is defined by a
$m$-linear smooth real function on the $k$-order jet manifold
(\ref{df145}), i.e., it is a bundle morphism
\mar{df150}\ben
&& \cE: J^k(\op\times_X^m F) \op\to_X F, \label{df150}\\
&& t'=\cE(x^\la,y^1_\La,\ldots,y^m_\La), \qquad |\La|\leq k,
\nonumber
\een
(cf. Definition \ref{df130}). Then by virtue of Definition
\ref{df125}, an $r$-order prolongation of the multidifferential
operator $\cE$ (\ref{df150}) is the morphism
\mar{df151}\ben
&& \cJ^r\cE: J^{r+k}(\op\times_X^m F) \op\to_X J^rF, \label{df151} \\
&& t'_\La\circ \cJ^r\cE=d_\La\cE, \qquad |\La|\leq r. \nonumber
\een

Accordingly, it follows from Definition \ref{df126} that the
infinite order jet prolongation of the multidifferential operator
$\cE$ (\ref{df150}) is the morphism
\mar{df152}\ben
&& \cJ^\infty\cE: J^\infty (\op\times_X^m F)\to J^\infty F,
\label{df152}\\
&& t'_\La\circ \cJ^\infty\cE=d_\La\cE, \qquad 0\leq|\La|. \nonumber
\een

\subsection{Star-product in a covariant form}

Let $X$ be a smooth manifold. We start with the notion of a
connection on $C^\infty(X)$-modules \cite{book00,book12}.

\begin{definition} \label{df401} \mar{df401} Let $P$ be a $C^\infty(X)$-module.
An $\mathbb R$-module morphism $\nabla$ $Q\to Q$ is called the
derivation of $Q$ if it satisfies the Leibniz rule
\be
\nabla(fq)=\dr(f)p + f\nabla(p), \qquad q\in Q, \qquad f\in
C^\infty(X),
\ee
where $\dr$ is a derivation of a real ring $C^\infty(X)$.
\end{definition}

Derivations of a  $C^\infty(X)$-module constitute a
$C^\infty(X)$-module that we denote $\gd Q$. In particular, a
derivation of a real ring $C^\infty(X)$ also is a derivation of
$C^\infty(X)$ as a $C^\infty(X)$-module. Let us recall that there
is one-to-one correspondence between the vector fields $\tau$ on
$X$ and the derivations
\be
\nabla_\tau (f)=\tau^\m\dr_\m f
\ee
of a real ring $C^\infty(X)$. this correspondence provides a
$C^\infty(X)$-module isomorphism $\gd C^\infty(X)=\cT_1(X)$.

\begin{definition} \label{df501} \mar{df501} A connection $\nabla$
on a $C^\infty(X)$-module $Q$ is defined to be a
$C^\infty(X)$-module monomorphism
\mar{df502}\beq
\nabla: \gd C^\infty(X)\to \gd Q \label{df502}
\eeq
which associates some derivation $\nabla_\tau$ of $Q$ to each
vector field $\tau$ on $X$.
\end{definition}

Given a fibre bundle $Y\to X$, let $J^\infty Y$ (\ref{df106}) be
the infinite order jet manifold  and $\cO^*_\infty$ (\ref{df505})
the $C^\infty(X)$-algebra  of exterior forms of finite jet order
on $X$. Let us consider the total differential $d_H$
(\ref{df506}). This is a $\mathbb R$-module morphism of
$\cO^*_\infty$ so that, for any vector field $\tau$ on $X$, it
defines a derivation
\mar{df507}\beq
\nabla_\tau(\phi)=\tau\rfloor d_H\f =\tau^\la d_\la \f, \qquad
f\in \cO^0_\infty, \label{df507}
\eeq
of a $C^\infty(X)$-module $\cO^0_\infty$ of smooth real functions
of finite jet order on $J^\infty Y$. In accordance with Definition
\ref{df501}, $d_H$ is a connection on a $C^\infty(X)$-module
$\cO^0_\infty$.

Let now $Y=F$ be the fibre bundle (\ref{df20}) coordinated by
$(x^\la,t)$ and $J^\infty F$ its infinite order jet manifold
(\ref{df105}). Let $C^\infty(J^\infty F)=\cO_r^0$ be a
$C^\infty(X)$-module (\ref{df45}) of smooth real functions of
finite jet order on $J^\infty F$ (Definition \ref{df250}). Then
the total differential $d_H$ where $d_\la$ are the total
derivatives (\ref{df34}) defines a canonical connection on
$C^\infty(J^rF)$ such that, for any vector field $\tau$, we have
the derivation (\ref{df507}):
\mar{df509}\beq
\nabla_\tau(\f)=\tau\rfloor d_H(\f) =\tau^\la d_\la \f, \qquad
\f\in C^\infty(J^rF), \label{df509}
\eeq
of $C^\infty(J^\infty F)$.

In accordance with Theorem \ref{df251}, a function  $f\in
C^\infty(J^\infty F)$ is a finite order differential operator on a
real ring $C^\infty(X)$. Its derivation $\nabla_\tau(\f)$
(\ref{df509}) also is an element of $C^\infty(J^\infty F)$ and,
consequently, a differential operator of finite jet order on
$C^\infty(X)$. It is represented as a composition of the following
morphisms
\mar{df510}\beq
J^\infty F\ar_X^{\cJ^\infty\f} J^\infty F \ar_X^{\nabla_\tau} F
\label{df510}
\eeq
where $\cJ^\infty\f$ is the infinite jet order prolongation
(\ref{df134}) of a differential operator $\f$ on $C^\infty(X)$ and
$\nabla_\tau=\tau^\la t_\la$. Due to the isomorphism (\ref{df25}),
this morphism is globally defined.

\begin{remark} \label{df511} \mar{df511}
The morphism $\nabla_\tau$ in the composition (\ref{df510})
requires first order jet prolongation of a differential operator
$\f$. In a general setting, if we consider an $r$-order
differential operator on $C^\infty(J^\infty F)$, they imply
$r$-order jet prolongation (\ref{df132}) of differential operators
$\f$ on $C^\infty(X)$.
\end{remark}

Turn now to multidifferential operators. Let us consider a product
bundle
\be
\op\times_X^m F\to X,
\ee
coordinated by $(x^\la, t^i)$, $i=1,\ldots,m$, where $t^i$ are
global coordinates possessing identity transition functions. Let
$\Pi^m$ be the corresponding infinite order jet manifold
(\ref{df147}). Let us consider a $C^\infty(X)$-module
$C^\infty(\Pi^m)$ of smooth real functions of finite jet order on
$\Pi^m$. In accordance with Theorem \ref{df148}, elements of
$C^\infty(\Pi^m)$ are $C^\infty(X)$-valued multidifferential
operators on $C^\infty(X)$.  A $C^\infty(X)$-module
$C^\infty(\Pi^m)$ is endowed with the canonical connection
\mar{df901}\beq
d_H=dx^\la\ot d_\la =dx^\la(\dr_\la +\op\sum_{0\leq|\La|,1\leq
i\leq m}t^i_{\la+\La}\dr_i^\La. \label{df901}
\eeq
Since $\Pi^m$ is the direct product, we also can consider the
connections
\mar{df512}\beq
d_H^i=dx^\la\ot d^i_\la =dx^\la(\dr_\la
+\op\sum_{0\leq|\La|}t^i_{\la+\La}\dr_i^\La) \label{df512}
\eeq
for every $i=1,\ldots,m$. Given a multivector field
\mar{df513}\beq
\vt =\frac{1}{m!}\vt^{\la_1\dots\la_m}(x) \dr_{\la_1}\w\cdots\w
\dr_{\la_m}, \label{df513}
\eeq
let us consider the composition of derivations
\mar{df514}\beq
\wh\vt= \frac{1}{m!}\vt^{\la_1\dots\la_m}(x)
d^1_{\la_1}\circ\cdots\circ d^m_{\la_m}, \label{df514}
\eeq
acting on a $C^\infty(X)$-module $C^\infty(\Pi^m)$. Owing to the
isomorphism (\ref{df25}), this composition is globally defined.
Let us call it a multiderivation.  We also have a smooth real
function
\mar{df814}\beq
\vt= \frac{1}{m!}\vt^{\la_1\dots\la_m}(x)
d^1_{\la_1}\circ\cdots\circ d^m_{\la_m}, \label{df814}
\eeq
Let us consider a function $vt_0=t^1\cdots t^m$ on $\Pi^m$ and its
multiderivation
\mar{df515}\beq
\wh\vt(\vt_0)=\frac{1}{m!}\vt^{\la_1\dots\la_m}(x)
t^1_{\la_1}\circ\cdots\circ t^m_{\la_m}. \label{df515}
\eeq
This is a smooth real function on $\Pi^m$ which provides a
multilinear differential operator on $C^\infty(X)$. One also
construct compositions of the multiderivations (\ref{df814}) that
however implies the corresponding jet prolongation of the operator
$\vt_0$ (Remark  \ref{df511}).

In particular, let consider the product $F\times F$ coordinated by
$(x^\la,t,t')$, and let  $C^\infty(\Pi^2)$ be a
$C^\infty(X)$-module of smooth real functions of finite jet order
on $\Pi^2$. It is endowed with the canonical connections $d_H^t$
and $d_H^{t'}$ (\ref{df512}). Let
\be
\ve=\frac12w^{\al\bt}(x)\dr_\al\w\dr_\bt
\ee
be a bivector field on $X$. It defines the corresponding
biderivation
\mar{df808}\beq
\wh \ve= \ve^{\la_1\dots\la_m} (x) d^t_{\la_1}\circ\cdots\circ
d^{t'}_{\la_m}, \label{df808}
\eeq
of a $C^\infty(X)$-module $C^\infty(\Pi^2)$. Given a deformation
parameter $h$, let us consider the composition of this
biderivation written in the compact form
\mar{df809}\beq
\exp[h\wh\ve]= \op\sum_{0\leq k} \frac{h^k}{k!}\op\circ^k\wh\ve
\cdots \wh\ve. \label{df809}
\eeq
Let this derivation act on a function $\ve_0=tt'\in
C^\infty(\Pi^2)$. Then the result
\mar{df810}\ben
&& t*t'=\exp[h\wh\ve](tt')= tt' + \frac12\ve(x){\al\bt}t_\al t'_\bt+
\label{df810}\\
&& \qquad \op\sum_{1\leq k}
\frac{h^k}{k!}\op\circ^k[\ve^{\al_1\bt_1}d_{\al_1}
d'_{\bt_1}]\circ\cdots\circ \cdots [\ve^{\al_k\bt_k}d_{\al_k}
d'_{\bt_k}](\ve(x){\al\bt}t_\al t'_\bt) \nonumber
\een
is a differential operator on $C^\infty(X)$. It can be treated as
a star-product.

A direct computation shows that the star-product obeys the
associativity condition.

\section{Appendix}

This Section summarizes the relevant material on fibre bundles and
jet manifolds \cite{book09,book00,book13,sau}.

All morphisms are smooth (i.e. of class $C^\infty$) and manifolds
are smooth real and finite-dimensional. A smooth manifold is
 conventionally assumed to be Hausdorff and second-countable. Consequently, it is
 locally compact and paracompact. Being
paracompact, a smooth manifold admits a partition of unity by
smooth real functions. Unless otherwise stated, manifolds are
assumed to be connected.

The standard symbols $\otimes$, $\vee$, $\wedge$ and $\rfloor$
stand for the tensor, symmetric, exterior and interior products,
respectively. By $C^\infty(Z)$ is denoted the ring of smooth real
functions on a manifold $Z$.

If $Z$ is a manifold $Z$, we denote by
\be
\pi_Z:TZ\to Z, \qquad \pi^*_Z:T^*Z\to Z
\ee
its tangent and cotangent bundles,  respectively. Given
coordinates $(z^\al)$ on $Z$, they are equipped with the holonomic
coordinates
\mar{df26,7}\ben
&& (z^\la,\dot z^\la), \qquad \dot z'^\la= \frac{\dr z'^\la}{\dr
z^\mu}\dot z^\m, \label{df26}\\
&&(z^\la,\dot z_\la), \qquad \dot z'_\la= \frac{\dr z'^\m}{\dr
z^\la}\dot z_\m, \label{df27}
\een
with respect to the holonomic frames $\{\dr_\la\}$ and  coframes
$\{dz^\la\}$ in the tangent and cotangent spaces to $Z$,
respectively. Any manifold morphism $f:Z\to Z'$ yields the tangent
morphism
\mar{ss46}\beq
Tf:TZ\to TZ', \qquad \dot z'^\la\circ Tf = \frac{\dr f^\la}{\dr
x^\m}\dot z^\m. \label{ss46}
\eeq

\subsection{Fibre bundles}

Let us consider manifold morphisms of maximal rank. They are
immersions and submersions. An injective immersion is a
submanifold. If an immersion $f$ is a homeomorphism onto $f(M)$
equipped with the relative topology from $N$, it is called the
imbedding. If $M\subset N$, its natural injection is denoted by
$i_M:M\to N$.

A surjective submersion
\mar{11f1}\beq
\pi :Y\to X, \qquad \di X=n>0, \label{11f1}
\eeq
is called the fibred manifold, i.e., the tangent morphism
$T\pi:TY\to TX$ is a surjection. One says that $Y$ is a total
space of the fibred manifold (\ref{11f1}), $X$ is its base, $\pi$
is a fibration,  and $Y_x=\pi^{-1}(x)$ is a fibre over $x\in X$.

\begin{theorem} \mar{11t1} \label{11t1} A surjection $Y\to X$
is a fired manifold iff a manifold $Y$ admits an atlas of fibred
coordinate charts $(U_Y; x^\la, y^i)$ such that $(x^\la)$ are
coordinates on $\pi(U_Y)\subset X$ and coordinate transition
functions read $x'^\la =f^\la(x^\m)$, $y'^i=f^i(x^\m,y^j)$.
\end{theorem}

The fibred manifold (\ref{11f1}) admits a local section. This is
an injection $s:U\to Y$ of an open subset $U\subset X$ such that
$\pi\circ s=\id U$.  If $U=X$, one calls $s$ the global section. A
global section need not exist.

\begin{theorem} \label{mos9} \mar{mos9}
A fibred manifold whose fibres are diffeomorphic to $\mathbb R^m$
always has a global section.
\end{theorem}

The fibred manifold $Y\to X$ (\ref{11f1}) is called the fibre
bundle if admits a fibred coordinate atlas $\{(\pi^{-1}(U_\xi);
x^\la, y^i)\}$ over a cover $\{\pi^{-1}(U_\iota)\}$ of $Y$ which
is the inverse image of a cover $\{U_\xi\}$ of $X$. In this case,
there exists a manifold $V$, called the typical fibre, such that
$Y$ is locally diffeomorphic to the splittings
\mar{mos02}\beq
\psi_\xi:\pi^{-1}(U_\xi) \to U_\xi\times V, \label{mos02}
\eeq
glued together by means of transition functions
\mar{mos271}\beq
\vr_{\xi\zeta}=\psi_\xi\circ\psi_\zeta^{-1}: U_\xi\cap
U_\zeta\times V \to  U_\xi\cap U_\zeta\times V \label{mos271}
\eeq
on overlaps $U_\xi\cap U_\zeta$. Restricted to a point $x\in X$,
trivialization morphisms $\psi_\xi$ (\ref{mos02}) and transition
functions $\vr_{\xi\zeta}$ (\ref{mos271}) define diffeomorphisms
of fibres
\mar{sp21,2}\ben
&&\psi_\xi(x): Y_x\to V, \qquad x\in U_\xi,\label{sp21}\\
&& \vr_{\xi\zeta}(x):V\to V, \qquad x\in U_\xi\cap U_\zeta. \label{sp22}
\een
 Trivialization charts $(U_\xi,
\psi_\xi)$ together with transition functions $\vr_{\xi\zeta}$
(\ref{mos271}) constitute a bundle atlas
\mar{sp5}\beq
\Psi = \{(U_\xi, \psi_\xi), \vr_{\xi\zeta}\} \label{sp5}
\eeq
of a fibre bundle $Y\to X$. Two bundle atlases are said to be
equivalent if their union also is a bundle atlas, i.e., there
exist transition functions between trivialization charts of
different atlases. A fibre bundle $Y\to X$ is uniquely defined by
a bundle atlas.

Without a loss of generality, we assume that a cover for a bundle
atlas of $Y\to X$ also is a cover for a manifold atlas of a base
$X$. Then, given the bundle atlas $\Psi$ (\ref{sp5}), a fibre
bundle $Y\to X$ is provided with the associated   bundle
coordinates
\be
x^\la(y)=(x^\la\circ \pi)(y), \qquad y^i(y)=(y^i\circ\psi_\xi)(y),
\qquad y\in \pi^{-1}(U_\xi),
\ee
where $x^\la$ are coordinates on $U_\xi\subset X$ and $y^i$,
called fibre coordinates, are coordinates on a typical fibre $V$.

A fibre bundle $Y\to X$ is said to be trivial if $Y$ is
diffeomorphic to a product $X\times V$.

\begin{theorem} \label{11t3} \mar{11t3} Any fibre
bundle over a contractible base is trivial.
\end{theorem}

A  bundle morphism of a fibre bundle $\pi:Y\to X$ to a fibre
bundle $\pi': Y'\to X'$, by definition, is a fibrewise morphism.
It is defined as a pair $(\Phi,f)$ of manifold morphisms which
form a commutative diagram
\be
\begin{array}{rcccl}
& Y &\ar^\Phi & Y'&\\
_\pi& \put(0,10){\vector(0,-1){20}} & & \put(0,10){\vector(0,-1){20}}&_{\pi'}\\
& X &\ar^f & X'&
\end{array}, \qquad \pi'\circ\Phi=f\circ\pi.
\ee

Bundle injections and surjections are called bundle monomorphisms
and epimorphisms, respectively. A bundle diffeomorphism is called
a bundle isomorphism, or a bundle automorphism if it is an
isomorphism to itself. For the sake of brevity, a bundle morphism
over $f=\id X$ is said to be a bundle morphism over $X$, and is
denoted by $Y\ar_XY'$. In particular, a bundle automorphism over
$X$ is called a vertical automorphism.

A bundle monomorphism $\Phi:Y\to Y'$ over $X$ is called a
subbundle of a fibre bundle $Y'\to X$ if $\Phi(Y)$ is a
submanifold of $Y'$.

Let us mention the following standard constructions of new fibre
bundles from the old ones.

$\bullet$ Given a fibre bundle $\pi:Y\to X$ and a manifold
morphism $f: X'\to X$, the pull-back of $Y$ by $f$ is called the
manifold
\mar{mos106}\beq
f^*Y =\{(x',y)\in X'\times Y \,: \,\, \pi(y) =f(x')\}
\label{mos106}
\eeq
together with a natural projection $(x',y)\to x'$. It is a fibre
bundle over $X'$ such that the fibre of $f^*Y$ over a point $x'\in
X'$ is that of $Y$ over a point $f(x')\in X$. There is the
canonical bundle morphism
\mar{mos81}\beq
f_Y:f^*Y\ni (x',y)|_{\pi(y) =f(x')} \op\to_f y\in Y. \label{mos81}
\eeq
Any section $s$ of a fibre bundle $Y\to X$ yields the  pull-back
section $f^*s(x')=(x',s(f(x'))$ of $f^*Y\to X'$.

$\bullet$ If $X'\subset X$ is a submanifold of $X$ and $i_{X'}$ is
the corresponding natural injection, then the pull-back bundle
$i_{X'}^*Y=Y|_{X'}$ is called the restriction of a fibre bundle
$Y$ to a submanifold $X'\subset X$. If $X'$ is an imbedded
submanifold, any section of the pull-back bundle $Y|_{X'}\to X'$
is the restriction to $X'$ of some section of $Y\to X$.

$\bullet$ Let $\pi:Y\to X$ and $\pi':Y'\to X$ be fibre bundles
over a base $X$. Their bundle product $Y\op\times_X Y'$ over $X$
is defined as the pull-back
\be
Y\op\times_X Y'=\pi^*Y'\quad \mathrm{or} \quad Y\op\times_X
Y'={\pi'}^*Y
\ee
together with its natural surjection onto $X$.  Fibres of the
bundle product $Y\op\times_X Y'$ are the Cartesian products
$Y_x\times Y'_x$ of fibres of fibre bundles $Y$ and $Y'$.

A vector bundle is a fibre bundle $Y\to X$ such that:

$\bullet$ its typical fibre $V$ and all fibres $Y_x=\pi^{-1}(x)$,
$x\in X$, are real finite-dimensional vector spaces;

$\bullet$ there is the bundle atlas $\Psi$ (\ref{sp5}) of $Y\to X$
whose trivialization morphisms $\psi_\xi$ (\ref{sp21}) and
transition functions $\vr_{\xi\zeta}$ (\ref{sp22}) are linear
isomorphisms.

Accordingly, a vector bundle is provided with linear bundle
coordinates $(y^i)$ possessing linear transition functions
$y'^i=A^i_j(x)y^j$. We have
\mar{trt}\beq
y=y^ie_i(\pi(y))=y^i \psi_\xi(\pi(y))^{-1}(e_i), \qquad \pi(y)\in
U_\xi, \label{trt}
\eeq
where $\{e_i\}$ is a fixed basis for the typical fibre $V$ of $Y$,
and $\{e_i(x)\}$ are the fibre bases (or the frames) for fibres
$Y_x$ of $Y$ associated to a bundle atlas $\Psi$.

By virtue of Theorem \ref{mos9}, any vector bundle has a global
section, e.g., the canonical global zero-valued section $\wh
0(x)=0$. Global sections of a vector bundle $Y\to X$ constitute a
$C^\infty(X)$-module $Y(X)$.

By a morphism of vector bundles is meant a linear bundle morphism,
 whose restriction to each fibre is a linear map.

There are the following particular constructions of new vector
bundles from the old ones.

$\bullet$ Let $Y\to X$ be a vector bundle with a typical fibre
$V$. By $Y^*\to X$ is denoted the dual vector bundle with the
typical fibre $V^*$ dual of $V$. The interior product of $Y$ and
$Y^*$ is defined as a fibred morphism
\be
\rfloor: Y\otimes Y^*\ar_X X\times \mathbb R.
\ee

$\bullet$ Let $Y\to X$ and $Y'\to X$ be vector bundles with
typical fibres $V$ and $V'$, respectively. Their Whitney sum
$Y\op\oplus_X Y'$ is a vector bundle over $X$ with the typical
fibre $V\oplus V'$.

$\bullet$ Given vector bundles $Y$ and $Y'$ over the same base
$X$, their tensor product $Y\ot Y'$ is a vector bundle over $X$
whose fibres are the tensor products of fibres of $Y$ and $Y'$.
Similarly, the exterior product $Y\w Y$ is defined. Accordingly,
\mar{spr880}\beq
\w Y=(X\times \mathbb R)\op\oplus_X Y\op\oplus_X \op\w^2
Y\op\oplus_X \cdots \op\oplus_X\op\w^m Y, \qquad m=\di
V,\label{spr880}
\eeq
is called the exterior bundle of $Y$.

The tangent bundle $TZ$ and the cotangent bundle $T^*Z$ of a
manifold $Z$ exemplify vector bundles. Given an atlas $\Psi_Z
=\{(U_\iota,\phi_\iota)\}$ of a manifold $Z$, the tangent bundle
is provided with the holonomic bundle atlas $\Psi_T =\{(U_\iota,
\psi_\iota = T\phi_\iota)\}$, where $T\phi_\iota$ is the tangent
morphism to $\f_\iota$. The associated linear bundle coordinates
are holonomic coordinates $(\dot z^\la)$ with respect to the
holonomic frames $\{\dr_\la\}$ in tangent spaces $T_zZ$.

A tensor product of tangent and cotangent bundles
\mar{sp20}\beq
T=(\op\ot^mTZ)\ot(\op\ot^kT^*Z), \qquad m,k\in \mathbb N,
\label{sp20}
\eeq
is called a tensor bundle, provided with holonomic bundle
coordinates $\dot x^{\al_1\cdots\al_m}_{\bt_1\cdots\bt_k}$
possessing transition functions
\be
\dot x'^{\al_1\cdots\al_m}_{\bt_1\cdots\bt_k}=\frac{\dr
x'^{\al_1}}{\dr x^{\m_1}}\cdots\frac{\dr x'^{\al_m}}{\dr
x^{\m_m}}\frac{\dr x^{\nu_1}}{\dr x'^{\bt_1}}\cdots\frac{\dr
x^{\nu_k}}{\dr x'^{\bt_k}} \dot
x^{\m_1\cdots\m_m}_{\nu_1\cdots\nu_k}.
\ee

Let $\pi_Y:TY\to Y$ be the tangent bundle of a fibre bundle $\pi:
Y\to X$. Given bundle coordinates $(x^\la,y^i)$ on $Y$, it is
equipped with the holonomic coordinates $(x^\la,y^i,\dot x^\la,
\dot y^i)$. The tangent bundle $TY\to Y$ has a subbundle $VY =
\Ker (T\pi)$,  which consists of the vectors tangent to fibres of
$Y$. It is called the vertical tangent bundle of $Y$,  and it is
provided with the holonomic coordinates $(x^\la,y^i,\dot y^i)$
with respect to the vertical frames $\{\dr_i\}$. Every bundle
morphism $\Phi: Y\to Y'$ yields a linear bundle morphism over
$\Phi$ of the vertical tangent bundles
\mar{ws538}\beq
V\Phi: VY\to VY', \qquad \dot y'^i\circ V\Phi=\frac{\dr
\Phi^i}{\dr y^j}\dot y^j. \label{ws538}
\eeq
It is called the vertical tangent morphism.

\subsection{Differential forms and multivector fields}

Vector fields on a manifold $Z$ are global sections of the tangent
bundle $TZ\to Z$. They are assembled into a real Lie algebra
$\cT_1(Z)$ with respect to the Lie bracket $[,]$.

A vector field $u$ on a fibre bundle $Y\to X$ is called
projectable if it projects onto a vector field on $X$, i.e., there
exists a vector field $\tau$ on $X$ such that $\tau\circ\pi=
T\pi\circ u$. Projectable vector fields take a coordinate form
\be
u=u^\la(x^\m) \dr_\la + u^i(x^\m,y^j) \dr_i, \qquad
\tau=u^\la\dr_\la.
\ee
A projectable vector field is called vertical if its projection
onto $X$ vanishes, i.e., it lives in the vertical tangent bundle
$VY$.

An exterior $r$-form on a manifold $Z$ is a section
\be
\f =\frac{1}{r!}\f_{\la_1\dots\la_r} dz^{\la_1}\w\cdots\w
dz^{\la_r}
\ee
of the exterior product $\op\w^r T^*Z\to Z$. Let $\cO^r(Z)$ denote
the vector space of exterior $r$-forms on a manifold $Z$. By
definition, $\cO^0(Z)=C^\infty(Z)$ is the ring of smooth real
functions on $Z$. All exterior forms on $Z$ constitute the
${\mathbb N}$-graded commutative algebra $\cO^*(Z)$ of global
sections of the exterior bundle $\w T^*Z$ (\ref{spr880}). This
algebra is provided with the exterior differential
\be
d\f= dz^\m\w \dr_\m\f=\frac{1}{r!} \dr_\m\f_{\la_1\ldots\la_r}
dz^\m\w dz^{\la_1}\w\cdots dz^{\la_r}.
\ee
This obeys the relations
\be
d\circ d=0, \qquad d(\f\w\si)= d(\f)\w \si +(-1)^{\nm\f}\f\w
d(\si),
\ee
where the symbol $|\f|$ stands for the form degree.

Given a manifold morphism $f:Z\to Z'$, any exterior $k$-form $\f$
on $Z'$ yields the pull-back exterior form $f^*\f$ on $Z$ given by
the condition
\be
f^*\f(v^1,\ldots,v^k)(z) = \f(Tf(v^1),\ldots,Tf(v^k))(f(z))
\ee
for an arbitrary collection of tangent vectors $v^1,\cdots, v^k\in
T_zZ$. The relations
\be
f^*(\f\w\si) =f^*\f\w f^*\si, \qquad  df^*\f =f^*(d\f)
\ee
hold. In particular, given a fibre bundle $\pi:Y\to X$, the
pull-back onto $Y$ of exterior forms on $X$ by $\pi$ provides the
monomorphism of graded commutative algebras $\cO^*(X)\to
\cO^*(Y)$. Elements of its range $\pi^*\cO^*(X)$ are called basic
forms. Exterior forms on $Y$ such that $u\rfloor\f=0$ for an
arbitrary vertical vector field $u$ on $Y$ are said to be
horizontal forms.

The interior product (or contraction) of a vector field $u$ and an
exterior $r$-form $\f$ on a manifold $Z$ is given by the
coordinate expression
\mar{d000}\beq
u\rfloor\f =  \frac1{(r-1)!} u^\mu \f_{\mu\la_2\ldots\la_r}
dz^{\la_2}\w\cdots \w dz^{\la_r}. \label{d000}
\eeq
It obeys the relations
\be
 \f(u_1,\ldots,u_r)=u_r\rfloor\cdots u_1\rfloor\f,\qquad
 u\rfloor(\f\w\si)= u\rfloor\f\w\si +(-1)^{\nm\f}\f\w
u\rfloor\si.
\ee

The Lie derivative of an exterior form $\f$ along a vector field
$u$ is
\be
\bL_u\f = u\rfloor d\f +d(u\rfloor\f), \qquad \bL_u(\f\w\si)=
\bL_u\f\w\si +\f\w\bL_u\si.
\ee
In particular, if $f$ is a function, then $\bL_u f =u(f)=u\rfloor
d f$.

Similarly to the graded commutative algebra of differential forms,
the graded commutative algebra of multivector fields on a manifold
$Z$ is introduced. A multivector field $\vt$ of degree $\nm\vt=r$
(or, simply, an $r$-vector field) on a manifold $Z$ is a section
\mar{cc6}\beq
\vt =\frac{1}{r!}\vt^{\la_1\dots\la_r} \dr_{\la_1}\w\cdots\w
\dr_{\la_r} \label{cc6}
\eeq
of the exterior product $\op\w^r TZ\to Z$.  Let $\cT_r(Z)$ denote
the vector space of $r$-vector fields on $Z$. By definition,
$\cT_0(Z)=C^\infty(Z)$. All multivector fields on a manifold $Z$
make up the $\mathbb N$-graded commutative  algebra $\cT_*(Z)$ of
global sections of the exterior bundle $\w TZ$ (\ref{spr880}) with
respect to the exterior product $\w$.

The graded commutative algebra $\cT_*(Z)$ is endowed with the
Schouten--Nijenhuis bracket
\mar{z30}\ben
&& [.,.]_\mathrm{SN}: \cT_r(M)\xx\cT_s(M) \to \cT_{r+s-1}(M), \label{z30} \\
&& [\vt,\up]_\mathrm{SN} =\vt\bullet\up
+(-1)^{rs}\up\bullet\vt, \nonumber \\
&& \vt\bullet\up = \frac{r}{r!s!}(\vt^{\mu\la_2\ldots\la_r}
\dr_\m\up^{\al_1\ldots\al_s}\dr_{\la_2}\w
\cdots\w\dr_{\la_r}\w\dr_{\al_1}\w\cdots\w\dr_{\al_s}). \nonumber
\een
This generalizes the Lie bracket of vector fields. The relations
\mar{gm6,7,8}\ben
&&[\vt,\up]_\mathrm{SN}=(-1)^{\nm\vt\nm\up}[\up,\vt]_\mathrm{SN}, \label{gm6}\\
&&[\nu,\vt\w\up]_\mathrm{SN} =[\nu,\vt]_\mathrm{SN}\w\up +(-1)^{(\nm\nu-1)\nm\vt}
\vt\w[\nu,\up]_\mathrm{SN},\label{gm7}\\
&&(-1)^{\nm\nu(\nm\up-1)}[\nu,[\vt,\up]_\mathrm{SN}]_\mathrm{SN} +
(-1)^{\nm\vt(\nm\nu-1)}[\vt,[\up,\nu]_\mathrm{SN}]_\mathrm{SN}
   \label{gm8}\\
&& \qquad + (-1)^{\nm\up(\nm\vt-1)}[\up,[\nu,\vt]_\mathrm{SN}]_\mathrm{SN}=0,\nonumber
\een
hold.

\begin{example} \label{df91} \mar{df91}
Let
\mar{df92}\beq
w=\frac12 w^{\mu\nu}\dr_\mu\w\dr_\nu \label{df92}
\eeq
be a bivector field. Given the multivector field $\vt$
(\ref{cc6}), the Schouten--Nijenhuis bracket $[w,\vt]_\mathrm{SN}$
(\ref{z30}) reads
\mar{df93}\ben
&& [w,\vt]_\mathrm{SN}= \label{df93} \\
&& \quad \left[\frac1{r!}w^{\mu\al}\dr_\mu\vt^{\la\la_2\ldots\la_r}+
\frac1{2(r-1)!}\vt^{\mu\la_2\ldots\la_r}\dr_\m
w^{\al\la}\right]\dr_\al\w\dr_\la\w\dr_{\la_2}\w\cdots\w\dr_{\la_r}.
\nonumber
\een
In particular,
\mar{df94}\beq
[w,w]_\mathrm{SN}=w^{\mu\la_1}\dr_\mu
w^{\la_2\la_3}\dr_{\la_1}\w\dr_{\la_2}\w\dr_{\la_3}. \label{df94}
\eeq
\end{example}

A generalization of the interior product (\ref{d000}) to
multivector fields is the left interior product
\be
\vt\rfloor\f=\f(\vt), \qquad \nm\vt\leq\nm\f, \qquad
\f\in\cO^*(Z), \qquad \vt\in \cT_*(Z),
\ee
of multivector fields and exterior forms. It is defined by the
equalities
\be
\f(u_1\w\cdots\w u_r)=\f(u_1,\ldots,u_r), \qquad \f\in\cO^*(Z),
\qquad u_i\in\cT_1(Z),
\ee
and obeys the relation
\be
\vt\rfloor \up\rfloor\f=(\up\w\vt)\rfloor\f=(-1)^{\nm\up\nm\vt}
\up\rfloor \vt\rfloor\f, \qquad \f\in\cO^*(Z), \qquad \vt,\up\in
\cT_*(Z).
\ee
If $\nm\f\leq\nm\vt$, the right interior product
\be
\vt\lfloor\f=\vt(\f), \qquad \f\in\cO^*(Z), \qquad \vt\in
\cT_*(Z),
\ee
of exterior forms and multivector fields is given by the
equalities
\mar{df11}\ben
&& \vt(\f_1,\ldots,\f_r)= \vt\lfloor\f_r\cdots\lfloor\f_1, \qquad \f_i\in
\cO^1(Z), \qquad \vt\in\cT_r(Z), \nonumber \\
   &&\vt\lfloor\f_i =
\frac1{(r-1)!}\vt^{\mu\al_2\ldots\al_r}\f_{i\mu}
\dr_{\al_2}\w\cdots\w  \dr_{\al_r}. \label{df11}
\een
It satisfies the relations
\be
&& (\vt\w\up)\lfloor \f=\vt\w (\up\lfloor\f) + (-1)^{\nm\up}(\vt\lfloor\f)
\w\up, \qquad \f\in\cO^1(Z), \\
&& \vt(\f\w\si)=\vt\lfloor\si\lfloor\f, \qquad \f,\si\in\cO^*(Z).
\ee

A tangent-valued $r$-form on a manifold $Z$ is a section
\mar{spr611}\beq
\phi = \frac{1}{r!}\phi_{\la_1\ldots\la_r}^\m dz^{\la_1}\w\cdots\w
dz^{\la_r}\ot\dr_\m \label{spr611}
\eeq
of the tensor bundle $\op\w^r T^*Z\ot TZ\to Z$.

A (regular) distribution on a manifold $Z$ is a subbundle $\bT$ of
the tangent bundle $TZ$ of $Z$. A vector field $u$ on $Z$ is said
to be subordinate to a distribution $\bT$ if it lives in $\bT$. A
distribution $\bT$ is called involutive if the Lie bracket of
$\bT$-subordinate vector fields also is subordinate to $\bT$. The
following local coordinates can be associated to an involutive
distribution  \cite{war}.

\begin{theorem}\label{c11.0} \mar{c11.0} Let $\bT$ be an involutive
$r$-dimensional distribution on manifold $Z$, $\di Z=k$. Every
point $z\in Z$ has an open neighborhood $U$ which is a domain of
an adapted coordinate chart $(z^1,\dots,z^k)$ such that,
restricted to $U$, the distribution $\bT$ and its annihilator
$\rA\bT$ are spanned by the local vector fields $\dr/\dr z^1,
\cdots,\dr/\dr z^r$ and the one-forms $dz^{r+1},\dots, dz^k$,
respectively.
\end{theorem}

A connected submanifold $N$ of a manifold $Z$ is said to be an
integral manifold of a distribution $\bT$ on $Z$ if $TN\subset
\bT$. Unless otherwise stated, by an integral manifold is meant an
integral manifold of dimension of $\bT$. An integral manifold is
called  maximal if no other integral manifold contains it. The
following is the classical theorem of Frobenius \cite{kob,war}.

\begin{theorem}\label{to.1}  \mar{to.1} Let $\bT$ be an
involutive distribution on a manifold $Z$. Through any point $z\in
Z$, there passes a unique maximal integral manifold of $\bT$, and
any integral manifold through $z$ is its open subset.
\end{theorem}

Maximal integral manifolds of an involutive distribution on a
manifold $Z$ are assembled into a (regular) foliation $\cF$ of
$Z$. It is defined as a partition of $Z$ into connected
$r$-dimensional submanifolds (leaves of a foliation) $F_\iota$,
$\iota\in I$, which possesses the following properties
\cite{rei,tam}. A foliated manifold $(Z,\cF)$ admits an adapted
coordinate atlas
\mar{spr850}\beq
\{(U_\xi;z^\la; z^i)\},\quad \la=1,\ldots,n-r, \qquad
i=1,\ldots,r, \label{spr850}
\eeq
such that transition functions of coordinates $z^\la$ are
independent of the remaining coordinates $z^i$ and, for each leaf
$F$ of a foliation $\cF$, the connected components of $F\cap
U_\xi$ are given by the equations $z^\la=$const. These connected
components and coordinates $(z^i)$ on them make up a coordinate
atlas of a leaf $F$.

\subsection{First order jet manifolds}

Given a fibre bundle $Y\to X$ with bundle coordinates
$(x^\la,y^i)$, let us consider the equivalence classes $j^1_xs$ of
its sections $s$ identified by their values $s^i(x)$ and values of
their derivatives $\dr_\mu s^i(x)$ at points $x\in X$. They are
called the first order jets of sections. The particular choice of
coordinates does not matter for this definition. The key point is
that the set  $J^1Y$ of first order jets $j^1_xs$, $x\in X$, is a
smooth manifold with respect to the adapted coordinates
$(x^\la,y^i,y_\la^i)$ such that
\mar{50}\beq
y_\la^i(j^1_xs)=\dr_\la s^i(x),\qquad {y'}^i_\la = \frac{\dr
x^\m}{\dr{x'}^\la}(\dr_\m +y^j_\m\dr_j)y'^i.\label{50}
\eeq
It is called the first order jet manifold of a fibre bundle $Y\to
X$ \cite{book,book13,sau}.

The jet manifold $J^1Y$ admits the natural fibrations
\mar{1.14}\beq
\pi^1:J^1Y\ni j^1_xs\to x\in X, \qquad \pi^1_0:J^1Y\ni j^1_xs\to
s(x)\in Y, \label{1.14}
\eeq
where the second one is an affine bundle.

There are the canonical imbeddings
\mar{18,24}\ben
&&\la_1:J^1Y\op\to_Y
T^*X \op\otimes_Y TY, \quad \la_1=dx^\la \otimes (\dr_\la +
y^i_\la \dr_i)=dx^\la\otimes
d_\la, \label{18}\\
&&\thh_1:J^1Y \op\to_Y T^*Y\op\otimes_Y VY, \quad
 \thh_1=(dy^i- y^i_\la dx^\la)\otimes \dr_i=\thh^i \otimes
\dr_i,\label{24}
\een
where $d_\la$ are total derivatives and $\thh^i$ are called
contact forms.  Identifying the jet manifold $J^1Y$ to its images
under the canonical morphisms (\ref{18}) and (\ref{24}), one can
represent jets $j^1_xs=(x^\la,y^i,y^i_\m)$ by tangent-valued forms
\mar{cc4}\beq
dx^\la \otimes (\dr_\la + y^i_\la \dr_i), \qquad (dy^i- y^i_\la
dx^\la)\otimes\dr_i. \label{cc4}
\eeq

Sections and morphisms of fibre bundles admit prolongations to jet
manifolds as follows.

$\bullet$ Every section $s$ of a fibre bundle $Y\to X$ has the jet
prolongation to the section
\be
(J^1s)(x): = j_x^1s, \qquad y_\la^i\circ J^1s= \dr_\la s^i(x),
\ee
of the jet bundle $J^1Y\to X$. A section $\ol s$ of the jet bundle
$J^1Y\to X$ is called holonomic or integrable if it is the jet
prolongation of some section of the fibre bundle $Y\to X$.

$\bullet$ Every bundle morphism $\Phi:Y\to Y'$ over a
diffeomorphism $f$ admits a jet prolongation to the bundle
morphism over $\Phi$ of the affine jet bundles
\be
J^1\Phi : J^1Y \ar_\Phi J^1Y',\qquad {y'}^i_\la\circ
J^1\Phi=\frac{\dr(f^{-1})^\m}{\dr x'^\la}d_\m\Phi^i.
\ee

$\bullet$ Every projectable vector field $u$ on a fibre bundle
$Y\to X$ has a jet prolongation to the projectable vector field
\mar{1.21}\beq
J^1u =u^\la\dr_\la + u^i\dr_i + (d_\la u^i - y_\m^i\dr_\la
u^\m)\dr_i^\la \label{1.21}
\eeq
on the affine jet bundle $J^1Y\to Y$.

\subsection{Higher and infinite order jets}

The notion of a first order jet manifolds is naturally extended to
higher order jets.

Let $Y\to X$ be a fibre bundle. Given its bundle coordinates
$(x^\la,y^i)$, a multi-index $\La$ of the length $\nm\La=r$
throughout denotes a collection of numbers $(\la_r...\la_1)$
modulo permutations. By $\La+\Si$ is meant a multi-index
$(\la_r\cdots\la_1\si_k\cdots\si_1)$. Let use the notation
\mar{df113}\beq
\dr_\La=\dr_{\la_r}\circ\cdots\circ\dr_{\la_1}. \label{df113}
\eeq

The  $r$-order jet manifold  $J^rY$  of sections of a fibre bundle
$Y\to X$ (or, simply, the jet manifold of $Y\to X$) is defined as
the disjoint union of the equivalence classes $j^r_xs$ of sections
$s$ of $Y$ such that different sections $s$ and $s'$ belong to the
same equivalence class $j^r_xs$ iff
\be
s^i(x)={s'}^i(x), \qquad\dr_\La s^i(x)=\dr_\La {s'}^i(x), \qquad
0< |\La| \leq r.
\ee
In brief, one can say that sections of $Y\to X$ are identified by
the $r+1$ terms of their Taylor series at points of $X$. The
particular choice of coordinates does not matter for this
definition. Given bundle coordinates $(x^\la,y^i)$ on a fibre
bundle
  $Y\to X$, the set $J^rY$ is endowed with an atlas
of the adapted coordinates
\mar{55.1,21}\ben
&& (x^\la, y^i_\La),\qquad   y^i_\La\circ s= \dr_\La s^i(x), \qquad
0\leq\nm\La \leq r, \label{55.1}\\
&& {y'}^i_{\la+\La}=\frac{\dr x^\m}{\dr'x^\la}d_\m y'^i_\La, \label{55.21}
\een
where the symbol $d_\la$ stands for the total derivative
\mar{5.32}\beq
d_\la = \dr_\la + \op\sum_{|\La|=0}^{r-1} y^i_{\La+\la}\dr_i^\La.
\label{5.32}
\eeq
Let us also use the notation
\mar{df114}\beq
d_\La=d_{\la_r}\circ\cdots\circ d_{\la_1}. \label{df114}
\eeq

The coordinates (\ref{55.1}) bring the set $J^rY$  into a smooth
manifold. They are compatible with the natural surjections
\be
\pi_l^r: J^rY\to J^lY, \quad r>l,
\ee
which form the composite bundle
\be
&& \pi^r: J^rY\op\ar^{\pi^r_{r-1}} J^{r-1}Y\op\ar^{\pi^{r-1}_{r-2}} \cdots
\op\ar^{\pi^1_0} Y\op\ar^\pi X, \\
&& \pi^k_h\circ\pi^r_k=\pi^r_h, \qquad \pi^h\circ\pi^r_h=\pi^r.
\ee
A glance at the transition functions (\ref{55.21}) when $\nm\La=r$
shows that the fibration $J^rY\to J^{r-1}Y$ is an affine bundle.

\begin{remark}
In order to introduce higher order jet manifolds, one can use the
construction of the repeated jet manifolds. Let us consider the
$r$-order jet manifold $J^rJ^kY$ of the jet bundle $J^kY\to X$. It
is coordinated by $(x^\m, y^i_{\Si\La})$, $|\La| \leq k$, $|\Si|
\leq r$. There is the canonical monomorphism
\mar{df112}\beq
\si_{rk}: J^{r+k}Y\to J^rJ^kY, \qquad y^i_{\Si\La}\circ \si_{rk}=
y^i_{\Si+\La}. \label{df112}
\eeq
\end{remark}

In the calculus in higher order jets, we have the $r$-order jet
prolongation functor such that, given fibre bundles $Y$ and $Y'$
over $X$, every bundle morphism $\Phi:Y\to Y'$ over a
diffeomorphism $f$ of $X$ admits the $r$-order jet prolongation to
a morphism of  $r$-order jet manifolds
\mar{5.152}\beq
J^r\Phi: J^rY\ni j^r_xs\to j^r_{f(x)}(\Phi\circ s\circ f^{-1}) \in
J^rY'. \label{5.152}
\eeq
The jet prolongation functor is exact. If $\Phi$ is an injection
(resp. a surjection), so is  $J^r\Phi$. It also preserves an
algebraic structure. In particular, if $Y\to X$ is a vector
bundle, so is $J^rY\to X$.

Every section $s$ of a fibre bundle $Y\to X$ admits the $r$-order
jet prolongation  to the section
\mar{df138}\beq
(J^rs)(x)= j^r_xs, \label{df138}
\eeq
called an integrable section, of the jet bundle $J^rY\to X$. Thus,
the jet prolongation functor yields an injection
\mar{df141}\beq
Y(X)\ni s\to J^rs\in J^rY(X) \label{df141}
\eeq
of a set $Y(X)$ of global sections of a fibre bundle $Y\to X$ onto
a subset of integrable sections (\ref{df138}) of a set of global
sections $J^rY(X)$ of the jet bundle $J^rY\to X$.

Every exterior form $\f$ on the jet manifold $J^kY$ gives rise to
the pull-back form $\pi^{k+i}_k{}^*\f$ on the jet manifold
$J^{k+i}Y$. Let $\cO_k^*=\cO^*(J^kY)$ denote the algebra of
exterior forms on the jet manifold $J^kY$. We  have the direct
system of differential graded algebras
\mar{5.7}\beq
\cO^*(X)\op\longrightarrow^{\pi^*} \cO^*(Y)
\op\longrightarrow^{\pi^1_0{}^*} \cO_1^*
\op\longrightarrow^{\pi^2_1{}^*} \cdots
\op\longrightarrow^{\pi^r_{r-1}{}^*}
  \cO_r^* \longrightarrow\cdots. \label{5.7}
\eeq

Higher order jet formalism provides the adequate formulation of
theory of (non-linear) differential operators
\cite{bryant,book,kras}.  Let $E\to X$ be a fibre bundle
coordinated by $(x^\la,v^A)$, $A=1,\ldots,m$.

\begin{definition}\label{ch538} \mar{ch538}
A $k$-order $E$-valued differential operator on a fibre bundle
$Y\to X$ is defined as a section $\Delta$ of the pull-back bundle
\mar{5.113}\beq
J^kY\op\times_X E \to J^kY. \label{5.113}
\eeq
\end{definition}

There is obvious one-to-one correspondence between the sections
$\Delta$ of the fibre bundle (\ref{5.113}) and the bundle
morphisms
\mar{df110}\beq
\Delta: J^kY\op\to_X E,  \qquad v^A\circ \Delta= \Delta^A(x^\la,
y^i, y^i_\la,\ldots, y^i_{\la_k\cdots \la_1}). \label{df110}
\eeq
Such a morphism also is called a $k$-order differential operator
on a fibre bundle $Y\to X$. It sends each section $s$ of $Y\to X$
onto the section
\mar{df139}\beq
(\Delta\circ J^ks)^A(x)= \Delta^A(x^\la, s^i(x), \dr_\la
s^i(x),\ldots, \dr_{\la_k}\cdots \dr_{\la_1}s^i(x)) \label{df139}
\eeq
of a fibre bundle $E\to X$. Therefore, there is the following
equivalent definition of differential operators on $Y$.

\begin{definition}\label{ch539} \mar{ch539}
Let $Y\to X$ and $E\to X$ be fibre bundles. The bundle morphism
$J^kY\to E$ (\ref{df110}) over $X$ is called the $E$-valued
$k$-order differential operator on $Y\to X$.
\end{definition}

Finite order jet manifolds make up the inverse system
\mar{5.10}\beq
X\op\longleftarrow^\pi Y\op\longleftarrow^{\pi^1_0}\cdots
\longleftarrow J^{r-1}Y \op\longleftarrow^{\pi^r_{r-1}}
J^rY\longleftarrow\cdots. \label{5.10}
\eeq
Its projective limit
\mar{df106}\beq
J^\infty Y=\op\lim^\lto J^rY, \label{df106}
\eeq
is defined as a set $J^\infty Y$, whose elements represent
$\infty$-jets $j^\infty_xs$ of local sections of $Y\to X$,
together with surjections
\mar{5.74}\beq
\pi^\infty: J^\infty Y\to X, \quad \pi^\infty_0: J^\infty Y\to Y,
\quad \quad \pi^\infty_k: J^\infty Y\to J^kY, \label{5.74}
\eeq
such that $\pi^\infty_r=\pi^k_r\circ \pi^\infty_k$ for any
admissible $k$ and $r<k$. Sections of $Y$  belong to the same jet
$j^\infty_xs$ iff their Taylor series at a point $x\in X$ coincide
with each other. Therefore, $J^\infty Y$ is called the infinite
order jet space.

The set $J^\infty Y$ is provided with the inverse limit topology.
This is the coarsest topology such that the surjections
(\ref{5.74}) are continuous. The base of open sets of this
topology in $J^\infty Y$ consists of the inverse images of open
subsets of $J^kY$, $k=0,\ldots$, under the mappings (\ref{5.74}).
This topology makes $J^\infty Y$ into a paracompact Fr\'echet
manifold. A bundle coordinate atlas $\{U,(x^\la,y^i)\}$ of $Y\to
X$ yields the manifold coordinate atlas
\mar{jet1}\beq
\{(\pi^\infty_0)^{-1}(U_Y), (x^\la, y^i_\La)\}, \qquad 0\leq|\La|,
\label{jet1}
\eeq
  of $J^\infty
Y$, together with the transition functions (\ref{55.21}) where
$d_\la$ denotes the total derivative
\mar{5.177}\beq
d_\la = \dr_\la + \op\sum_{0\leq|\La|}
y^i_{\la+\La}\dr_i^\La.\label{5.177}
\eeq

Though $J^\infty Y$ fails to be a smooth manifold, one can
introduce the differential calculus on $J^\infty Y$ as follows.

Let us consider the direct system (\ref{5.7}) of $\mathbb
R$-modules $\cO^*_k$ of exterior forms on finite order jet
manifolds $J^kY$. Its direct limit
\mar{df505}\beq
\cO^*_\infty=\op\lim^\to \cO^*_k \label{df505}
\eeq
together with monomorphisms
\be
\pi^{\infty*}_r:\cO^*_k\to \cO^*_\infty
\ee
is an $C^\infty(X)$-module which consists of all the exterior
forms on finite order jet manifolds module pull-back
identification. The operations of the exterior product $\w$ and
the exterior differential $d$ also have the direct limits on
$\cO_\infty^*$. They provide $\cO_\infty^*$ with the structure of
the differential graded algebra
\mar{5.13} \beq
0\to \mathbb R\longrightarrow
\cO^0_\infty\op\longrightarrow^d\cO^1_\infty\op\longrightarrow^d
\cdots, \label{5.13}
  \eeq
where $\cO^m_\infty$  are the direct limits of the direct systems
\mar{df38}\beq
\cO^m(X)\op\longrightarrow^{\pi^*} \cO^m_0
\op\longrightarrow^{\pi^1_0{}^*} \cO_1^m \longrightarrow \cdots
  \cO_r^m \op\longrightarrow^{\pi^{r+1}_r{}^*} \cO_{r+1}^m \longrightarrow
\cdots \label{df38}
\eeq
of real vector spaces $\cO_r^m$ of exterior $m$-forms on $r$-order
jet manifolds $J^rY$. We agree to call elements of $\cO^*_\infty$
the exterior forms on the infinite order jet space $J^\infty Y$.

Given a manifold coordinate atlas (\ref{jet1}) of $J^\infty Y$,
elements of the direct limit $\cO^*_\infty$ can be written in the
coordinate form as exterior forms on finite order jet manifolds.
The basic one-forms $dx^\la$ and the contact forms
$\thh^i_\La=dy^i_\la-y^i_{\la+\La} dx^\la$ make up the set of
local generators of the $\cO^0_\infty$-algebra $\cO_\infty^*$.

Accordingly, the exterior differential on $\cO_\infty^*$ is split
into the sum $d=d_H+d_V$ of the horizontal (total) and vertical
differentials such that
\mar{df506}\ben
&&  d_H(\f)=
dx^\la\w d_\la(\f), \qquad
d_V(\f)=\thh^i_\La \w \dr^\La_i\f, \qquad \f\in\cO^*_\infty, \label{df506}\\
&& d_H\circ d_H=0, \qquad d_V\circ d_V=0, \qquad d_V\circ d_H +d_H\circ
d_V=0.\nonumber
\een

\subsection{Hochschild cohomology}

This and next Sections summarize the relevant basics on homology
and cohomology of algebraic systems \cite{fuks,book05,mcl,massey}.

Let $\cK$ be a commutative ring and $\cA$ a $\cK$-ring which need
not be commutative. Hochschild cohomology is the cohomology of a
$\cK$-ring $\cA$ with coefficients in a  $\cA$-bimodule.
Throughout this Section, by $\ot$ is meant the tensor product of
modules over $\cK$.

Let $B_k(\cA)$, $k\in\mathbb N_+$, be an $\cA$-bimodule whose
basis is the tensor product of $\cK$-modules $\op\ot^k\cA$, i.e.,
it consists of elements $[a_1\ot\cdots\ot a_k]$, $a_i\in \cA$. The
$B_0(\cA)$ is defined as an $\cA$-bimodule of rank 1 whose basis
element is denoted by $[\;\;]$. One considers the chain complex
$B_*(\cA)$ with respect to boundary operators defined as the
$\cA$-bimodule morphisms
\mar{spr95}\ben
&& \dr_{k>0}[a_1\ot\cdots\ot a_k]=a_1[a_2\ot\cdots\ot a_k] +\label{spr95}\\
&& \qquad  \op\sum_{j=1}^{k-1} (-1)^j
[a_1\ot\cdots\ot a_ja_{j+1}\ot\cdots \ot a_k]
   + (-1)^k[a_1\ot\cdots \ot a_{k-1}]a_k.
\nonumber
\een
In particular, we have
\be
\dr_0: B_0(\cA)\ni a[\;\;]a'\to aa'\in \cA, \qquad
\dr_1(a[a_1]a')=aa_1[\;\;]a'-a[\;\;]a_1a'.
\ee

This chain complex admits the homotopy operator given by the right
$\cA$-module morphisms
\mar{spr100}\beq
\mathbf{h}:a[a_1\ot\cdots \ot a_k] \to [a\ot a_1\ot\cdots \ot
a_k], \qquad k\in\mathbb N. \label{spr100}
\eeq
For instance, we have $\mathbf{h}_0(a[\;\;]a')=[a]a'$. It follows
that the chain complex $B_*(\cA)$ is acyclic. The corresponding
exact sequence reads
\mar{spr101}\beq
0\lla \cA \lla^{\dr_0} B_0(\cA)\lla\cdots B_k(\cA)\lla^{\dr_{k+1}}
\cdots. \label{spr101}
\eeq
It is readily observed that the homotopy operator $\mathbf{h}$
(\ref{spr100}), completed by the map
\be
h: \cA\ni a\to [\;\;]a \in B_0(\cA),
\ee
is the homotopy operator for the chain complex (\ref{spr101}).
Therefore, this chain complex is exact at all terms $B_k(\cA)$,
$k\in\mathbb N$. Moreover, it is exact at $\cA$ since $\cA$ has a
unit element and, therefore, $\dr_0$ is a $\cK$-module
epimorphism. Hence, the chain complex (\ref{spr101}) is a
resolution of the ring $\cA$ by $\cA$-bimodules.

\begin{remark}
Given the canonical monomorphism of $\cK\to\cA$, one also
constructs the chain complex of $\cA$-bimodules
\mar{spr109}\beq
C_0(\cA)=B_0(\cA), \qquad C_{k>0}(\cA)=\op\ot^k (\cA/\cK)
\label{spr109}
\eeq
whose boundary operators take the form (\ref{spr95}). The chain
complexes $B_*(\cA)$ and $C_*(\cA)$ are proved to be homotopic
\cite{mcl}.
\end{remark}

Let us turn now to Hochschild cohomology. Let $Q$ be an
$\cA$-bimodule. Given the chain complex $B_*(\cA)$, let us
consider the cochain complex
\mar{ws50}\beq
0\to B^0(\cA,Q)\ar^{\dl^0}B^1(\cA,Q)\ar\cdots
B^k(\cA,Q)\ar^{\dl^k} \cdots, \label{ws50}
\eeq
whose terms are the $\cA$-bimodules
\be
B^k(\cA,Q)=\hm_\cK(B_k(\cA),Q).
\ee
It is called the Hochschild complex. Their elements can be seen as
$Q$-valued $\cK$-multilinear functions $f^k(a_1,\ldots, a_k)$ on
$\cA$. The coboundary operators read
\mar{spr103}\ben
&&(\dl^kf^k)(a_1,\ldots,a_{k+1})=f^k(\dr_{k+1}[a_1\ot
\cdots \ot a_{k+1}])= \label{spr103}\\
&& \qquad a_1f^k(a_2,\ldots,a_{k+1}) + \op\sum_j(-1)^jf^k
(a_1,\ldots,a_ja_{j+1},\ldots,a_{k+1}) \nonumber\\
&& \qquad  + (-1)^{k+1}f^k(a_1,\ldots,a_k)a_{k+1}, \qquad k\in \mathbb
N.
\nonumber
\een
In particular, the module $B^0(\cA,Q)$ is isomorphic to $Q$ via
the association
\be
Q\ni q\to f_q^0\in B^0(\cA,Q), \qquad f_q^0([\;\;])=q.
\ee
For instance, we have
\mar{spr104-6}\ben
&& \dl^0 f^0_q(a)=aq-qa, \qquad a\in \cA, \label{spr104}\\
&& \dl^1f^1(a_1,a_2)=a_1f^1(a_2)-f^1(a_1a_2) +f^1(a_1)a_2, \label{spr105}\\
&& \dl^2 f^2(a_1,a_2,a_3)=  a_1f^2(a_2,a_3) - f^2(a_1a_2,a_3)
+  \label{spr106}\\
&& \qquad f^2(a_1,a_2a_3) - f^2(a_1,a_2)a_3, \qquad
a_1,a_2,a_3\in\cA. \nonumber
\een

\begin{definition} \label{df16} \mar{df16}
Cohomology $H^*(\cA,Q)$ of the complex $B^*(\cA,Q)$ (\ref{ws50})
is called the Hochschild cohomology of a $\cK$-ring $\cA$ with
coefficients in an $\cA$-module $Q$.
\end{definition}

The Hochschild complex $B^*(\cA,Q)$ (\ref{ws50}) is  functorial in
$Q$, i.e., every $\cA$-bimodule morphism $Q\to P$ yields a cochain
$\cK$-module morphism $B^*(\cA,Q)\to B^*(\cA,P)$ and,
consequently, a homomorphism of Hochschild cohomology groups
$H^*(\cA,Q)\to H^*(\cA,P)$ \cite{gerst}. In particular, an
$\cA$-bimodule morphism $\Phi:Q\to P$ is called $\cK$-allowable if
there exists a $\cK$-module morphism $\la:P\to Q$ such that
$\Phi\circ\la\circ\Phi=\Phi$. This condition is always satisfied
if $\cK$ is a field.

The Hochschild cohomology modules of low degrees have the
following natural interpretation.

(i) Since $H^0(\cA,Q)=\Ker \dl^0=Q$, it follows from the
expression (\ref{spr104}) that $H^0(\cA,Q)$ is isomorphic to the
center $\cZ_Q$ of the $\cA$-bimodule $Q$.

(ii) A glance at the expression (\ref{spr105}) shows that any
normalized one-cocycle $f^1$ is a $Q$-valued derivation of the
$\cK$-ring $\cA$ and {\it vice versa}. In particular,
one-coboundaries $\dl^0f^0_q$ are inner $Q$-valued derivations up
to the sign minus. Therefore, $H^1(\cA,Q)$ is sometimes called the
module of outer derivations of $\cA$.

In particular, let $Q$ be $\cA$ itself seen as an $\cA$-bimodule.
In this case, the Hochschild complex $B^*(\cA,\cA)$ is provided
with the following two products \cite{gerst}.

(i) The first one is the cup-product
\mar{ws71}\beq
f\smile
f'(a_1,\ldots,a_{m+n})=f(a_1,\ldots,a_m)f'(a_{m+1},\ldots,a_{m+n}),
\label{ws71}
\eeq
which obeys the equality
\be
\dl^{m+n}(f\smile f')=\dl^m f\smile f'+(-1)^m f\smile \dl^n f'.
\ee
Hence, it induces the cup-product
\mar{ws75}\beq
[f]\smile [f']=[f\smile f'] \label{ws75}
\eeq
of the Hochschild cohomology classes $[f],[f']\in H^*(\cA,\cA)$.
This product makes $H^*(\cA,\cA)$ into a graded commutative
algebra, i.e., the relation
\be
[f]\smile [f']=(-1)^{|f||f'|}[f']\smile [f]
\ee
holds, where $|f|$ denotes the degree of cohomology classes.

(ii) The composition product of $f\in B^m(\cA,\cA)$ and $f'\in
B^n(\cA,\cA)$ is defined by the formula
\mar{ws70}\ben
&& f\circ f'(a_1,\ldots,a_{m+n-1})= \label{ws70}\\
&&\quad \op\sum_{i=1}^m (-1)^{(i-1)(n-1)}
f(a_1,\ldots,a_{i-1},f'(a_i,\ldots, a_{n+i-1}), a_{n+i},\ldots,
a_{n+m-1}). \nonumber
\een
It obeys the relation
\be
\dl(f\circ f')=(-1)^{n-1}\dl f\circ f' +f\circ\dl f'
+(-1)^n(f'\smile f- (-1)^{mn}f\smile f').
\ee
It is easily justified that, if $f$ and $f'$ are cocycles, the
bracket
\mar{ws73}\beq
[f,f']^\circ=f\circ f'-(-1)^{(m-1)(n-1)}f'\circ f \label{ws73}
\eeq
is also a cocycle. Consequently, the bracket (\ref{ws73}) induces
the bracket of the Hochschild cohomology classes
\mar{ws72}\beq
[[f],[f']]^\circ =[[f,f']^\circ], \label{ws72}
\eeq
called the Gerstenhaber bracket. This bracket makes $H^*(\cA,\cA)$
into a graded Lie algebra where elements $[f]\in H^m(\cA,\cA)$ are
provided with the graded degree $|f|-1=m-1$.

The cup-product (\ref{ws75}) and the Gerstenhaber bracket
(\ref{ws72}) in the Hochschild cohomology $H^*(\cA,\cA)$ satisfy
the relations making $H^*(\cA,\cA)$ into a Gerstenhaber algebra.

\subsection{Chevalley--Eilenberg cohomology of Lie algebras}

One associates the Chevalley--Eilenberg complex to a Lie algebra.

In this Section, $\cG$ denotes a Lie algebra (not necessarily
finite-dimensional) over a commutative ring $\cK$.

Let $\cG$ act on a $\cK$-module $P$ on the left by endomorphisms
\be
&& \cG\times P\ni (\ve,p)\to \ve p\in P, \\
&& [\ve,\ve']p=(\ve\circ
\ve'-\ve'\circ \ve)p, \qquad \ve,\ve'\in\cG.
\ee
One says that $P$ is a $\cG$-module.  A $\cK$-multilinear
skew-symmetric map
\be
c^k:\op\times^k\cG\to P
\ee
is called a $P$-valued $k$-cochain on the Lie algebra $\cG$. These
cochains form a $\cG$-module $C^k[\cG;P]$. Let us put
$C^0[\cG;P]=P$. We obtain the cochain complex
\mar{spr997}\beq
0\to P\ar^{\dl^0} C^1[\cG;P]\ar^{\dl^1} \cdots C^k[\cG;P]
\ar^{\dl^k} \cdots, \label{spr997}
\eeq
with respect to the Chevalley--Eilenberg coboundary operators
\mar{spr132}\ben
&& \dl^kc^k (\ve_0,\ldots,\ve_k)=\op\sum_{i=0}^k(-1)^i\ve_ic^k(\ve_0,\ldots,
\wh\ve_i, \ldots, \ve_k)+ \label{spr132}\\
&& \qquad \op\sum_{1\leq i<j\leq k}
(-1)^{i+j}c^k([\ve_i,\ve_j], \ve_0,\ldots, \wh\ve_i, \ldots,
\wh\ve_j,\ldots, \ve_k), \nonumber
\een
where the caret $\,\wh{}\,$ denotes omission \cite{fuks}. The
complex (\ref{spr997}) is called the Chevalley--Eilenberg complex
with coefficients in a module $P$. It is finite if the Lie algebra
$\cG$ is finite-dimensional.

For instance,
\mar{spr133,4}\ben
&& \dl^0p(\ve_0)=\ve_0p, \label{spr133}\\
&& \dl^1c^1(\ve_0,\ve_1)=\ve_0c^1(\ve_1)-\ve_1c^1(\ve_0) -
c^1([\ve_0,\ve_1]). \label{spr134}
\een

Cohomology $H^*(\cG;P)$ of the complex $C^*[\cG;P]$ is called the
Chevalley--Eilenberg cohomology of the Lie algebra $\cG$ with
coefficients in the module $P$.

In particular, let $P=\cG$  be regarded as a $\cG$-module with
respect to the adjoint representation
\be
\ve:\ve'\to [\ve,\ve']\qquad, \ve,\ve'\in\cG.
\ee
We abbreviate with $C^*[\cG]$ the Chevalley--Eilenberg complex of
$\cG$-valued cochains on $\cG$.

\begin{definition} \label{df15} \mar{df15}
Cohomology $H^*(\cG)$ of this complex is called the
Chevalley--Eilenberg cohomology or, simply, the cohomology of a
Lie algebra $\cG$.
\end{definition}

In particular, $C^0[\cG]=\cG$, while $C^1[\cG]$ consists of
endomorphisms of the Lie algebra $\cG$. Accordingly, the
coboundary operators (\ref{spr133}) and (\ref{spr134}) read
\mar{spr135,6}\ben
&&\dl^0\ve(\ve_0)=[\ve_0,\ve], \label{spr135} \\
&& \dl^1 c^1(\ve_0,\ve_1)=[\ve_0,c^1(\ve_1)]-[\ve_1,c^1(\ve_0)] -
c^1([\ve_0,\ve_1]). \label{spr136}
\een
A glance at the expression (\ref{spr136}) shows that a one-cocycle
$c^1$ on $\cG$ obeys the relation
\be
c^1([\ve_0,\ve_1])=[c^1(\ve_0),\ve_1]+ [\ve_0,c^1(\ve_1)]
\ee
and, thus, it is a derivation of the Lie algebra $\cG$.
Accordingly, any one-coboundary (\ref{spr135}) is an inner
derivation of $\cG$ up to the sign minus. Therefore, one can think
of the cohomology $H^1(\cG)$ as being the set of outer derivations
of $\cG$. One can show that the cohomology $H^2(\cG)$ indexes
non-equivalent infinitesimal deformations of the Lie algebra $\cG$
\cite{fuks}.


\begin{thebibliography}{ederf}

\bibitem{abr} R.Abraham  and J.Marsden, \emph{Foundations of Mechanics}
(Benjamin / Cummings Publ. Comp., London, 1978).

\bibitem{bayen} F.Bayen, M.Flato, C.Fronsdal, A.Lichnerowicz and
D.Sternheimer, Deformation thjeory and quantization I, II,
\emph{Ann. Phys.} \textbf{111} (1978) 61.

\bibitem{bert97} M.Bertelson, M.Cahen and S.Gutt, Equivalence of
star products, \emph{Class. Quant. Grav.} \textbf{14} (1997) A93.

\bibitem{bou} F.Bourgeois and M.Cahen, Variational principle for symplectic
connections, \emph{J. Geom. Phys.} \textbf{30} (1999) 233.

\bibitem{bryant} R.Bryant, S.Chern, R.Gardner, H.Goldschmidt and P.Griffiths,
\emph{Exterior Differential Systems} (Springer, Berlin, 1991).


\bibitem{cat02} A.Cattaneo, G.Fedler and L.Tomassini, From global
to local deformation quantization of Poisson manifolds, \emph{Duke
Math. J.} \textbf{115} (2002) 329.

\bibitem{cat04} A.Cattaneo, Formality and star-product, \emph{E-print arXiv}: math.QA/0403135.

\bibitem{delign} P.Deligne, D\'eformations de l'alg\'ebre des
fonctions d'une vari\`et\`e symplectique: Comparaison entre
Fedosov et De Wilde, Lecomte, \emph{Selecta Math. (New Series)}
\textbf{1} (1995) 667.


\bibitem{wild} M.DeWilde and P.Lecomte, Existence of
star-products and of formal deformations of the Poisson Lie
algebra of arbitrary symplectic manifold, \emph{Lett. Math. Phys.}
\textbf{7} (1983) 487.

\bibitem{dito} G.Dito and D.Sternheimer, Deformation quantization:
genesis, developments and metamorphoses, \emph{E-print arXiv}:
math.QA/0201168.

\bibitem{dolg} V.Dolgushev, Covariant and equivariant formality
theorem, \emph{E-print arXiv}: math.QA/0307212.


\bibitem{fed} B.Fedosov, A simple geometrical construction of
deformation quantization, \emph{J. Diff. Geom} \textbf{40} (1994)
213.

\bibitem{fed1} B.Fedosov, \emph{Deformation Quantization and
Index Theory}, Mathematical Topics, \textbf{9} (Akademie Verlag,
Berlin, 1996).

\bibitem{fial} A.Fialowski and M.Penkava, Deformation theory of
infinity algebras, \emph{Journal of Algebra} \textbf{255} (2002)
59.

\bibitem{fuks} D.Fuks, \emph{Cohomology of Infinite-Dimensional Lie
Algebras} (Consultants Bureau, New York, 1986).


\bibitem{gelf} I.Gelfand, V.Retakh and M.Shubin, Fedosov manifolds, \emph{Adv. Math.} \textbf{136} (1998) 104.

\bibitem{gerst} M.Gerstenhaber and S.Schack, Algebraic cohomology and
deformation theory, In: \emph{Deformation Theory of Algebras and
Structures and Applications}, NATO ASI Ser.C \textbf{247} (Kluwer,
Dordrecht, 1988) p.11.

\bibitem{gerst92} M.Gerstenhaber and S.Schack, Algebras,
bialgebras, quantum groups, and algebraic deformations,
\emph{Contemp. Math.} \textbf{134} (1992) 51.

\bibitem{book} G.Giachetta, L.Mangiarotti and G.Sardanashvily, \emph{New
Lagrangian and Hamiltonian Methods in Field Theory} (World
Scientific, Singapore, 1997).


\bibitem{book05} G.Giachetta, L.Mangiarotti and G.Sardanashvily,
\emph{Geometric and Algebraic Topological Methods in Quantum
Mechanics} (World Scientific, Singapore, 2005).

\bibitem{book09} G.Giachetta, L.Mangiarotti and G.Sardanashvily,
\emph{Advanced Classical Field Theory} (World Scientific,
Singapore, 2009).

\bibitem{gutt} S.Gutt and J.Rawnsley, Equivalence of star
products on a symplectic manifold; an introduction to Deligne's \v
Cech cohomology classes, \emph{J. Geom. Phys.} \textbf{29} (1999)
347.

\bibitem{han} F.Hansen, Quantum mechanics in phase space, \emph{Rep. Math.
Phys.} \textbf{19} (1984) 361.

\bibitem{hoch} G.Hochschild, B.Kostant and A.Rosenberg,
Differential forms on regular affine algebras, \emph{Trans. Am.
Math. Soc.} \textbf{10} (1962) 383.


\bibitem{kamm} J.Kammerer, Analysis of the Moyal product in flat space, \emph{J. Math. Phys.} \textbf{27} (1986) 529.

\bibitem{kob} S.Kobayashi and K.Nomizu, \emph{Foundations of Differential
Geometry, Vol.1,2.}  (Interscience Publ., N.Y., 1963, 1969).

\bibitem{kol} I.Kol\'a\v{r},  P.Michor and J.Slov\'ak, \emph{Natural Operations
in Differential Geometry} (Springer-Verlag, Berlin, 1993).

\bibitem{konts0} M.Kontsevich, Deformation quantization of Poisson
manifolds I, \emph{E-print arXiv}: q-alg/979040.

\bibitem{konts} M.Kontsevich, Deformation quantization of
algebraic varieties, \emph{Lett. Math. Phys.} \textbf{65} (2001)
271.

\bibitem{konts2} M.Kontsevich, Deformation quantization of Poisson
manifolds, \emph{Lett. Math. Phys.} \textbf{66} (2003) 157.

\bibitem{kras} I.Krasil'shchik, V.Lychagin and A.Vinogradov, \emph{Geometry of
Jet Spaces and Nonlinear Partial Differential Equations} (Gordon
and Breach, Glasgow, 1985).


\bibitem{libe} P.Libermann and C-M.Marle, \emph{Symplectic Geometry and
Analytical Mechanics} (D.Reidel Publishing Company, Dordrecht,
1987).

\bibitem{lich82} A.Lichnerowicz, Vari\'et\'es de Poisson et
feuilletages, \emph{Ann. Fac. Toulouse} (1982) 195.

\bibitem{mcl} S.Mac Lane, \emph{Homology} (Springer, Berlin, 1967).

\bibitem{book00} L.Mangiarotti and G.Sardanashvily, \emph{Connections in
Classical and Quantum Field Theory} (World Scientific, Singapore,
2000).

\bibitem{marl} C.-M.Marle, The Schouten--Nijenhuis bracket and
interior products, \emph{J. Geom. Phys.} \textbf{23} (1997) 350.

\bibitem{masm} M.Masmoudi, Tangential formal deformations of
the Poisson bracket and tangential star product on a regular
Poisson manifold, \emph{J. Geom. Phys.} \textbf{9} (1992) 155.

\bibitem{massey} W.Massey, \emph{Homology and Cohomology Theory} (Marcel Dekker,
Inc., New York, 1978).

\bibitem{nad99} F.Nadaud, On continuous and differential
Hochschild cohomology, \emph{Lett. Math. Phys.} \textbf{47} (1999)
85.

\bibitem{nest} R.Nest and B.Tsygan, Algebraic index theorem
for families, \emph{Adv. Math.} \textbf{113} (1995) 151.

\bibitem{omori} H.Omori, Y.Maeda and A.Yoshioka, Existence of a
closed star product, \emph{Lett. Math. Phys.} \textbf{26} (1992)
285.


\bibitem{pinc1} G.Pinczon, On the equivalence between continuous and differential
deformation theories, \emph{Lett. Math. Phys.} \textbf{39} (1997)
143.

\bibitem{pelaum} M.Pelaum, On continuous Hochschild homology and
cohomology groups, \emph{Lett. Math. Phys.} \textbf{44} (1998) 43.


\bibitem{rei} B.Reinhart, \emph{Differential Geometry and Foliations}
(Springer, Berlin, 1983).

\bibitem{rob} A.Robertson and W.Robertson, \emph{Topological Vector Spaces}
(Cambridge Univ. Press., Cambridge, 1973).

\bibitem{book12} G.Sardanashvily, \emph{Lectures on Differential Geometry of Modules and Rings.
Application to Quantum Theory} (Lambert Academic Publishing,
Saarbrucken, 2012); \emph{arXiv}: 0910.0092.

\bibitem{book13} G.Sardanashvily, \emph{Advanced Differential Geometry for Theoreticians.
Fiber bundles, jet manifolds and Lagrangian theory} (Lambert
Academic Publishing, Saarbrucken, 2013); \emph{arXiv}: 0908.1886.

\bibitem{sau} D.Saunders, \emph{The Geometry of Jet Bundles}
(Cambridge Univ. Press, Cambridge, 1989).

\bibitem{tam} I.Tamura, \emph{Topology of Foliations: An Introduction},
Transl. Math. Monographs \textbf{97} (AMS, Providence, 1992).


\bibitem{vais85} I.Vaisman, Symplectic curvature tensors, \emph{Monatsh.
Mathematik} \textbf{100} (1985) 299.

\bibitem{vais} I.Vaisman, \emph{Lectures on the Geometry of Poisson Manifolds}
(Birkh\"auser Verlag, Basel, 1994).

\bibitem{vey} J.Vey, D\'eformation du crochet de Poisson sur
une vari\'et\'e symplectique, \emph{Comment. Math. Helv.}
\textbf{50} (1975) 421.

\bibitem{war} F.Warner, \emph{Foundations of Differential Manifolds and Lie
Groups} (Springer, Berlin, 1983).

\bibitem{wein} A.Weinstein, The local structure of Poisson manifolds,
\emph{J. Diff. Geom.} \textbf{18} (1983) 523.

\bibitem{xu98} P.Xu, Fedosov $*$-products and quantum momentum
maps, \emph{Commun. Math. Phys.} \textbf{197} (1998) 167.



\end{thebibliography}
\end{document}